\documentclass[twocolumn,numbering,showpacs,nofootinbib,superscriptaddress]{revtex4-1}

\usepackage{graphicx,color}
\usepackage{latexsym}
\usepackage{amsmath,amssymb}        
\usepackage[draft=false, colorlinks=true, linkcolor = red, citecolor = blue]{hyperref}
\usepackage{mathrsfs}
\usepackage{comment}
\usepackage{soul}
\usepackage{mathrsfs}
\DeclareSymbolFontAlphabet{\mathrsfs}{rsfs}
\DeclareMathAlphabet{\mathcal}{OMS}{cmsy}{m}{n}

\usepackage{ulem}

\begin{document}


\title{Scalar field dark matter as an alternative explanation for the anisotropic distribution of satellite galaxies}


\author{Jordi Sol\'is-L\'opez} 
\affiliation{Departamento de F\'{i}sica, Centro de Investigaci\'on y de Estudios Avanzados del IPN, A.P. 14-740, 07000 CDMX, M\'exico}

\author{Francisco S. Guzm\'an}
\affiliation{Laboratorio de Inteligencia Artificial y Superc\'omputo,
	      Instituto de F\'{\i}sica y Matem\'{a}ticas, Universidad
              Michoacana de San Nicol\'as de Hidalgo. Edificio C-3, Cd.
              Universitaria, 58040 Morelia, Michoac\'{a}n,
              M\'{e}xico}
              
\author{Tonatiuh Matos} 
\affiliation{Departamento de F\'{i}sica, Centro de Investigaci\'on y de Estudios Avanzados del IPN, A.P. 14-740, 07000 CDMX, M\'exico}

\author{Victor H. Robles}
\affiliation{Yale Center for Astronomy \& Astrophysics, Physics Department, P.O. Box 208120, New Haven, Connecticut 06520, USA}
\affiliation{Physics Department, Yale University, P.O. Box 208120, New Haven, Connecticut 06520, USA}
\author{L. Arturo Ure\~na-L\'opez}
\affiliation{Departamento de F\'isica, DCI, Campus Le\'on, Universidad de Guanajuato, 37150, Le\'on, Guanajuato, M\'exico}
              

\date{\today}


\begin{abstract}
In recent years, the scalar field dark matter (SFDM), also called ultralight bosonic dark matter, has received considerable attention due to the number of problems it might help to solve. Among these are the cusp-core problem and the abundance of small structures of the standard cold dark matter (CDM) model. In this paper we show that multi-state solutions of the low energy and weak gravitational field limit of field equations, interpreted as galactic halo density profiles, can provide a possible explanation to the anisotropic distribution of satellite galaxies observed in the Milky Way, M31 and Centaurus A, where satellites trajectories seem to concentrate on planes close to the poles of the galaxies instead of following homogeneously distributed trajectories. The core hypothesis is that multi-state solutions of the equations describing the dynamics of this dark matter candidate, namely, the Gross-Pitaevskii-Poisson equations, with monopolar and dipolar contributions, can possibly explain the anisotropy of satellite trajectories. { In order to construct a proof of concept, we study the trajectories of a number of test particles traveling on top of the gravitational potential due to a multi-state halo with modes (1,0,0)+(2,1,0). The result is that particles accumulate asymptotically in time on planes passing close to the poles. Satellite galaxies are not test particles but interpreted as such, our results indicate that in the asymptotic time their trajectories do not distribute isotropically, instead they prefer to have orbital poles accumulating near the equatorial plane of the multistate halo. The concentration of orbital poles depends on whether the potential is monopolar or dipolar dominated.}
\end{abstract}


\pacs{keywords: dark matter -- Milky Way-- halos}

\maketitle

\section{Introduction}
\label{sec:introduction}
The standard model of cosmology $\Lambda$CDM assumes that dark matter (DM) is made of particles that interact only gravitationally and have low velocity dispersion, generically known as cold dark matter (CDM). The advances on the knowledge of structure formation at large scales, the distribution of DM and its comparison with the observed distribution of structures, are in great degree possible due to CDM simulations \cite{Navarro:1996gj,Springel:2005nw, Sawala:2015cdf}. These simulations reveal that a bottom-up hierarchical structure formation model holds and that galactic and cluster structures clump to end with a self-similar shape \cite{Navarro:1996gj,wang19}.

However, there are observations that are still puzzling to understand in the CDM model \cite{bullock17}. These include the well-known cusp-core problem \citep{santos19} and 
the excess of substructure abundance \cite{Klypin_1999,Moore_1999}. 
More recently, it has been suggested that satellite galaxies around the Milky Way (MW) accumulate near the galactic poles in the  vast polar structure (VPOS)\cite{Pawlowski:2019bar} (see also Fig. 6 in \cite{Pawlowski:2013}). The motion of satellite galaxies around M31 shows to be nonisotropic \cite{Conn_2013,ibata2013vast}. Moreover, CDM simulations predict that satellites hosted by the Milky Way rarely display the observed coherence of satellite positions and orbits \cite{marcel18}.

Among the 50 satellites in the Local Group, 43 are contained in four different planes \cite{Shaya:2013,Pawlowski:2013}, which is inconsistent with the isotropy predicted by simulations based on CDM. Some possible explanations within the CDM frame are still plausible. For instance, that there are more satellites outside of the VPOS that are still too faint to be detected.
Other possibilities to explain the plane of satellites is that interactions between gas and radiation might affect the isotropy of the final distribution of satellites; or that the Milky Way and M31 are atypical galaxies in which this unexpected coherent distribution of dwarfs happens. 
Nevertheless, recently it has been reported that a set of 31 satellites in the constellation of Centaurus interacts gravitationally with the elliptical galaxy Centaurus A and displays a similar anisotropic alignment \cite{Muller:2018hks}.
The probability of finding such anisotropic satellite distribution in CDM simulations is less than 0.5\% \cite{Muller:2018hks}. Noticing that there are now three galaxies, Cen A, Milky Way and M31, of at least two different types, all showing this anisotropy, indicates the possible  need of an explanation to the satellite distributions based on different grounds. 


One possible mechanism to break the isotropy of satellites could come from the scalar field dark matter (SFDM) model. This alternative model to CDM started at the end of the last century (see for example \cite{Sin:1992bg,Matos:1992qx}). The first systematic study of this model began in \cite{Matos:1998vk}, since then, this same model has appeared under various names, like fuzzy dark matter \cite{Hu:2000ke}, quintessential dark matter \cite{Arbey:2001qi}, wave dark matter \cite{Bray:2010fc} and more recently as ultralight dark matter \cite{Hui:2016ltb}.
The main idea of the model assumes the dark matter is an ultralight spin-0 boson such that its associated de Broglie wavelength will be of galactic scales, leading to quantum-like phenomena at the scale of galaxies and larger.
The first time the cosmology of this model was analyzed was in  \cite{Matos:2000ss}, where the boson mass is a parameter that determines the cutoff scale of the mass power spectrum. It was then determined that the boson mass had to be ultralight of order $\mu \sim 10^{-22}\text{eV}/c^2$. With this mass, the model could mimic the behavior of CDM model at cosmological scales, having the same mass power spectrum and the CMB spectrum~\cite{Matos:2000ss,Hlozek:2014lca}.

The first essential difference with CDM was that the SFDM model has a natural cutoff of the mass power spectrum at small scales established by the boson mass, which would be more consistent with the estimated amount of satellites, unlike CDM which predicts a higher power at those scales \cite{Matos:2000ss}. 
Another main difference is the central density distribution in SFDM halos. As shown in previous works \cite{robles12a,robles12,Bernal:2005hw,robles18}, SFDM halos have inner flat density profiles (cores) instead of cuspy density profiles as in $\Lambda$CDM. 
Later on, in \cite{Martinez-Medina:2014hca} numerical simulations with gas containing a SFDM halo show appropriate rotation curves for LSB galaxies, in \cite{Martinez-Medina:2015jra} the spiral arms were generated resembling real galaxies and \cite{medina15} show that dwarf spheroidal galaxies are also well modeled. Recent cosmological simulations of the first galaxies in SFDM \cite{mocz19,mocz19b} reveal early-forming cores in the dark matter, gas and stellar components. Surviving structures show the expected central density cores  \cite{Schive:2014dra,Mocz:2017wlg,hopkins19} resulting from the Heisenberg uncertainty principle preventing cusps at galactic scales. 

In this paper, we explore the possibility that multistate configurations of bosonic equilibrium configurations, { considered as DM halos, could explain the observations of VPOS due to the anisotropy of the different density modes. If a halo is a multistate configuration, there will be a preferential direction where the mass concentration is higher, or equivalently local minimums of the gravitational potential that will influence the trajectories of particles and structures within the halo}. Consequently particles traveling around will distribute in a nonisotropic manner, which {might} eventually explain the coherent motion of satellite galaxies in the MW, Andromeda and Centaurus A. 

The paper is organized as follows. In Sec. II we describe the multi-state configurations used to test our idea. In Sec. III we show the test particles analysis on which we base motion of satellites. In Sec. IV we present a set of consistency checks of our methods. Finally in Sec. V we discuss our results. 

\section{Multistate configurations}
\label{sec:model}

To explain the anisotropic distribution of satellites we first assume the gravitational potential of the host-galaxy halo is dominated by SFDM, whereas  satellites are assumed to behave as test particles orbiting around the halo. Second, we assume the low energy and weak field regimes to hold, which is valid in the galactic  scale regime. Third, under these conditions the resulting scalar field is the order parameter of the Gross-Pitaevskii-Poisson system (GPP)  that rules the dynamics of a condensate of bosons in coherent states $\Psi_{nlm}$, whose equations of motion are \cite{Nuevo}
\begin{subequations}
\label{eq:gpp}
\begin{eqnarray}
i\hbar \frac{\partial \Psi_{nlm}}{\partial t}&=&-\frac{\hbar^2}{2\mu}\nabla^2\Psi_{nlm}+\mu V\Psi_{nlm}, \label{eq:gpp-a} \\
\nabla^2 V&=& 4\pi G \hat{\mu}^2 c^2 \sum_{nlm}|\Psi_{nlm}|^2 \label{eq:gpp-b},
\end{eqnarray}
\end{subequations}
\noindent where $\mu$ is the boson mass, $c$  the speed of light, $\hbar$  the reduced Planck constant, $G$  the gravitational constant {and $\hat{\mu}$ is defined by  $\hat{\mu}\equiv \mu c/\hbar$}. 

{Notice that the Compton length of the boson particle is precisely $L_C =\hat{\mu}^{-1}$, which establishes the typical length scale of the  configurations. It is useful to fix units in terms of a mass scale, for which we set $\hat{\mu}^{-1} = 0.1 \mathrm{pc} (10^{-22} \mathrm{eV}/\mu c^2) $. Likewise, the typical timescale is given by $T_C = L_C/c = (\hat{\mu} c)^{-1}$, and then $T_C = 3 \mathrm{yr} (10^{-22} \mathrm{eV}/\mu c^2)$.}

The GPP system (\ref{eq:gpp}) is invariant under the scaling property $ \{ t, \mathbf{x}, \Psi, V \} \rightarrow \{ \lambda^{-2} t, \lambda^{-1} \mathbf{x}, \lambda^2 \Psi, \lambda^2 V \}$, for any real parameter $\lambda$ (see e.g. \cite{GuzmanUrena2004}). {Using this scaling property, we find that appropriate galactic-size scales are obtained assuming a boson mass of $\mu c^2 = 10^{-25} \text{eV}$, corresponding to $\hat{\mu}= 15.65/\text{kpc}$, and $\lambda \simeq 10^{-3}$. Unless explicitly stated, hereafter the latter will be our fiducial values for the physical examples studied below.\footnote{Such mass scale $\mu c^2 = 10^{-25} \text{eV}$ is below the expected value for SFDM models, see for instance~\cite{Davies_2020} and references therein. However, one cannot discard the existence of more SFDM, or axion, species as playing different roles at galactic scales~\cite{grin2019gravitational,Arvanitaki_2010,luu2020multiple}. In this respect, the results reported here should be understood as a guidance for more involved studies of galaxy dynamics under the SFDM hypothesis.}}

We now search for stationary solutions of multistate wave functions of this system of equations. For this we assume the wave function has the following expression in spherical coordinates
\begin{equation}
\Psi_{nlm}(t,r,\theta,\varphi)=\frac{e^{-i\gamma_{nlm}t}}{\sqrt{4\pi G}} r^l \psi_{nlm}(r)Y_{lm}(\theta,\varphi) ,
\label{eq:stationarywfs}
\end{equation}
\noindent with $\gamma_{nlm}$ an eigenfrequency obtained from a well-posed eigenvalue problem as described in \cite{Nuevo}, and $n=1,2,...$, $l=1,2,...,n-1$ and $m=-1,-l+1,...,l-1,l$. {If we use $V \to V c^2$, $r \to \hat{\mu}^{-1} r$ and $\gamma_{nlm} \to \hat{\mu} c \, \gamma_{nlm}$, together with the scaling property mentioned above, then Eqs.~\eqref{eq:gpp} becomes a fully dimensionless, scale-free, system for the quantities $\psi_{nlm}$ and $V$.}

Following the recipe in \cite{Nuevo}, we construct stationary solutions with the spherical and first dipolar contributions.  That is, we solve for the combination of states $\Psi_{100}$ together with $\Psi_{210}$. The reasons for this choice are, first, that this is the simplest non-spherically symmetric multi-state configuration after the spherical equilibrium configuration with (100)-mode only, second, the resulting two blobs associated to the dipolar (210)-mode are expected to pull test particles toward the poles and third, in \cite{Nuevo} a possible mechanism for the formation of such structures has been envisioned.

For our purposes, we use two workhorse examples. The first one with a dominant dipolar contribution such that the mass ratio between the spherical and dipolar masses is $M_{100}/M_{210}=0.36$, and has  eigenfrequencies $\gamma_{100}=1.8$ and $\gamma_{210}=1.42$. The second one has a dominant spherical contribution  with $M_{100}/M_{210}=3$, and eigenfrequencies $\gamma_{100}=0.84$, $\gamma_{210}=0.54$. {The mass scale of the configurations is $M_C = c^2/(G \hat{\mu}) = 10^{12} \, M_\odot (10^{-22} \mathrm{eV}/\mu c^2)$, and then the physical mass of each multipolar contribution of a configuration is obtained from $\lambda M_{100} M_C$ and $\lambda M_{210} M_C$. In the same manner, the physical eigenfrequencies are given by $\lambda^2 \gamma_{100} /T_C$ and $\lambda^2 \gamma_{210} /T_C$.}

\begin{figure*}[tp!]
\centering
\includegraphics[width=0.4\textwidth]{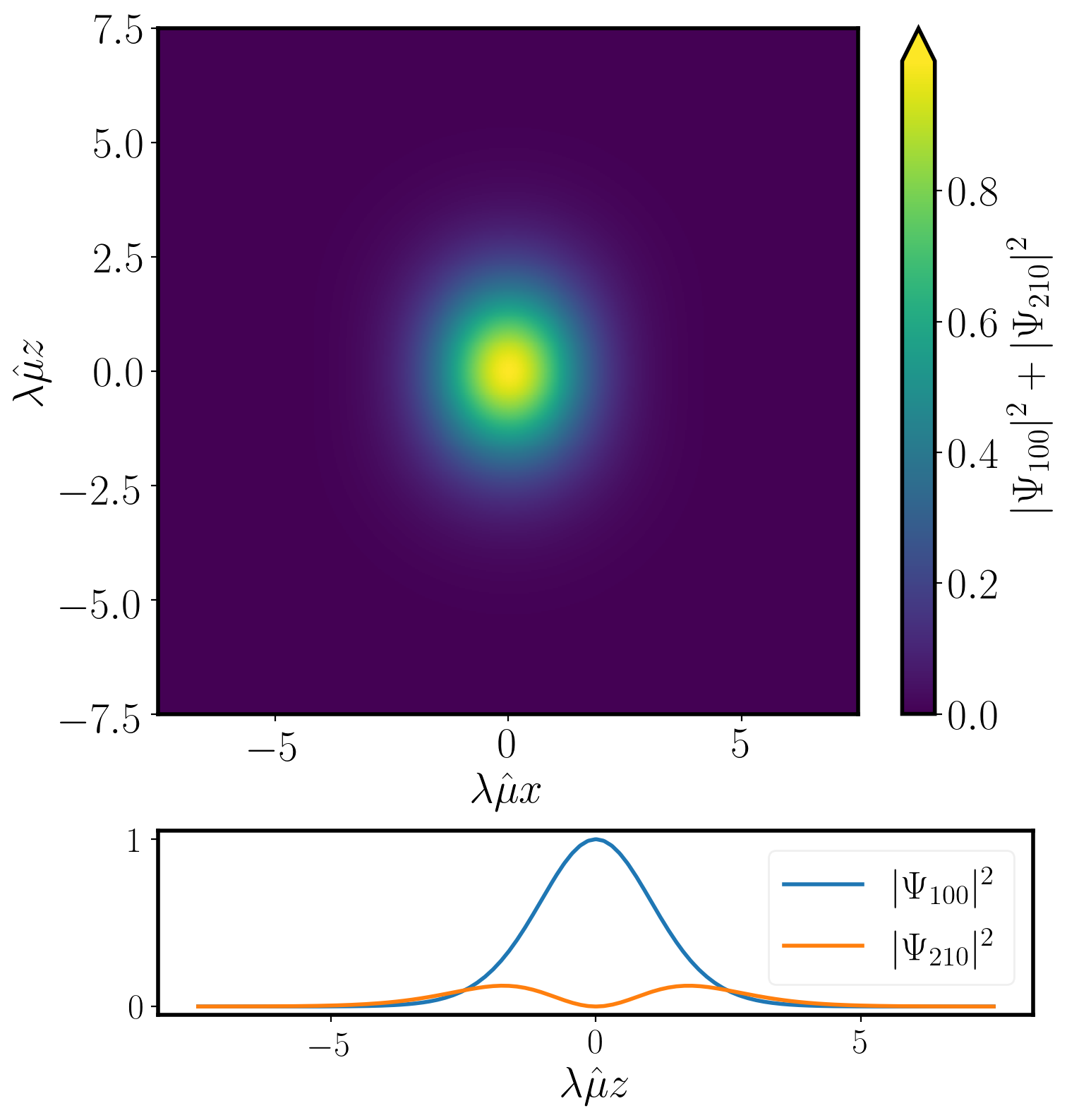}
\includegraphics[width=0.4\textwidth]{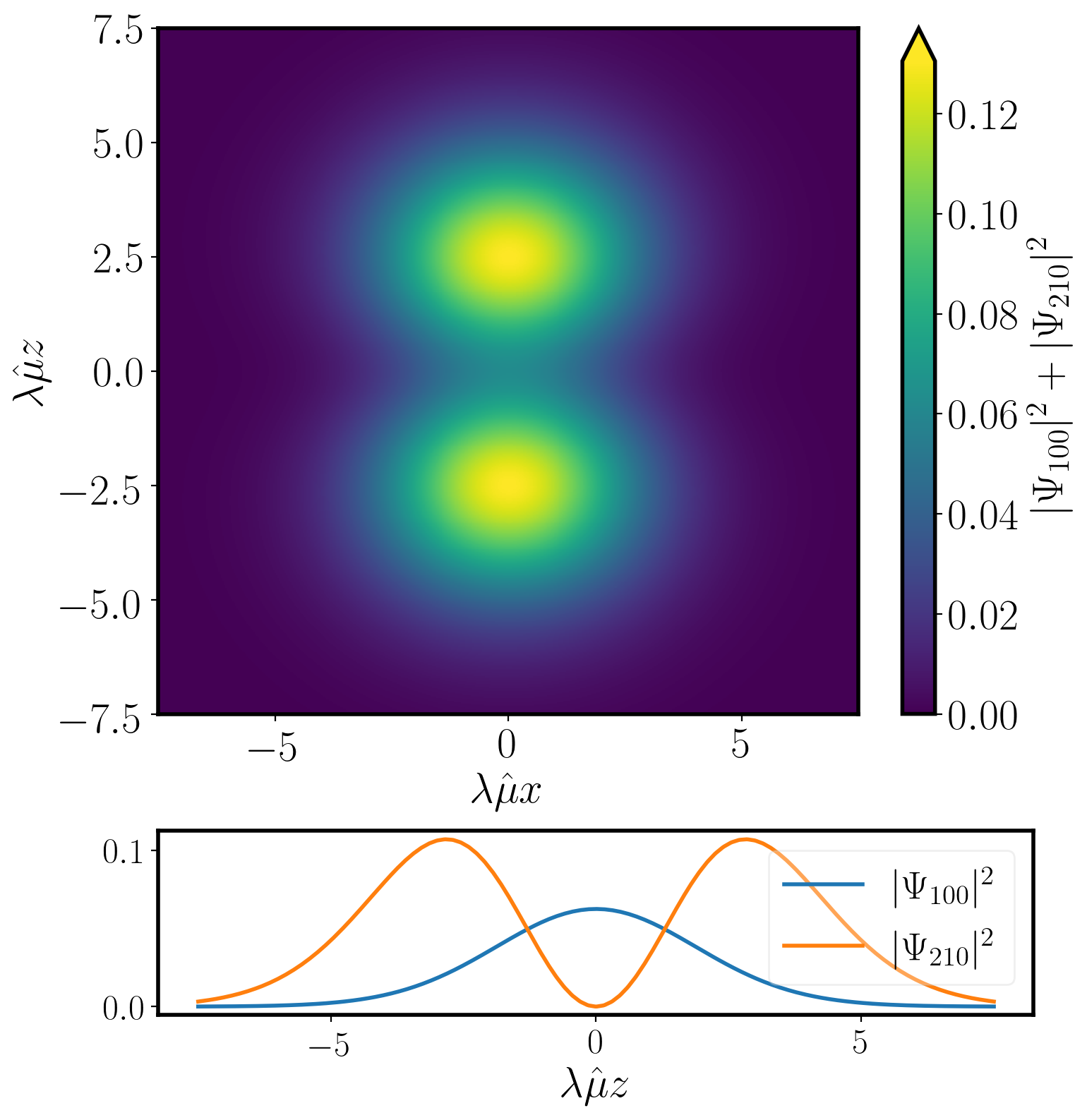}
\caption{Density distribution of two different multi-state configurations with states 100 and 210. The configuration on the left is a monopole-dominant configuration with mass ratio between states of $M_{100}/M_{210} = 3$. The one on the right is a dipole-dominating configuration such that the mass ratio between the spherical and dipolar contributions is $M_{100}/M_{210} = 0.36$. 
At the top of each panel we show the projection of the multistate configuration density $|\Psi_{100}|^2 + |\Psi_{210 }|^2$ on the $xz-$plane in dimensionless units, whereas at the bottom we show the projection of $|\Psi_{100}|^2$ and $|\Psi_{210}|^2$ along the $z-$axis.}
\label{fig:SchrPois_0}
\end{figure*}

The mass density of these two configurations is shown in Fig. \ref{fig:SchrPois_0}, {in terms of dimensionless and scale-free quantities.} These configurations will play the role of DM halos, whose density distribution generates the dominant gravitational potential of a galaxy. { The main difference between the two is the notorious presence of the dipole blobs in the dipole-dominating configuration. We will illustrate different scenarios using these two configurations, that we call monopole-dominating and dipole-dominating configurations respectively.}

\section{Analysis of test particle trajectories \label{sec:particles}}

{ Satellite  galaxies are not test particles, however the motion of test particles can indicate where the nondominating structures within a galactic potential accumulate with a certain likelihood. This is why we now study the motion of test particles within the gravitational potential $V(\mathbf{x})$ of Eq. (\ref{eq:gpp}) for the multistate host halo.} 

{ To study the effects of the multi-state  configuration density on test particles, for instance, where these particles would accumulate in the asymptotic time}, we integrate the trajectories of  $10^5$ particles. {The initial positions of the test particles are randomly chosen from a uniform distribution over the radius interval $(0, R=4]$.} 
{Similarly, initial velocities are random in direction  with velocity magnitude randomly chosen from a uniform distribution over the interval $(0, v_{max}=a v_{\rm{esc}}]$,} where $v_{\rm{esc}}$ is the escape velocity of a particle { at radius $R = 4$.} We analyze the system for $a=1/4,1/2,3/4$ and 1, however use the case $a=1/2$ to illustrate our analysis in detail in what follows, and show a comparison of results for the other values of $a$ in Appendix \ref{app:otherAss}. 

\begin{figure}[htp]
\centering
\includegraphics[width=0.21\textwidth]{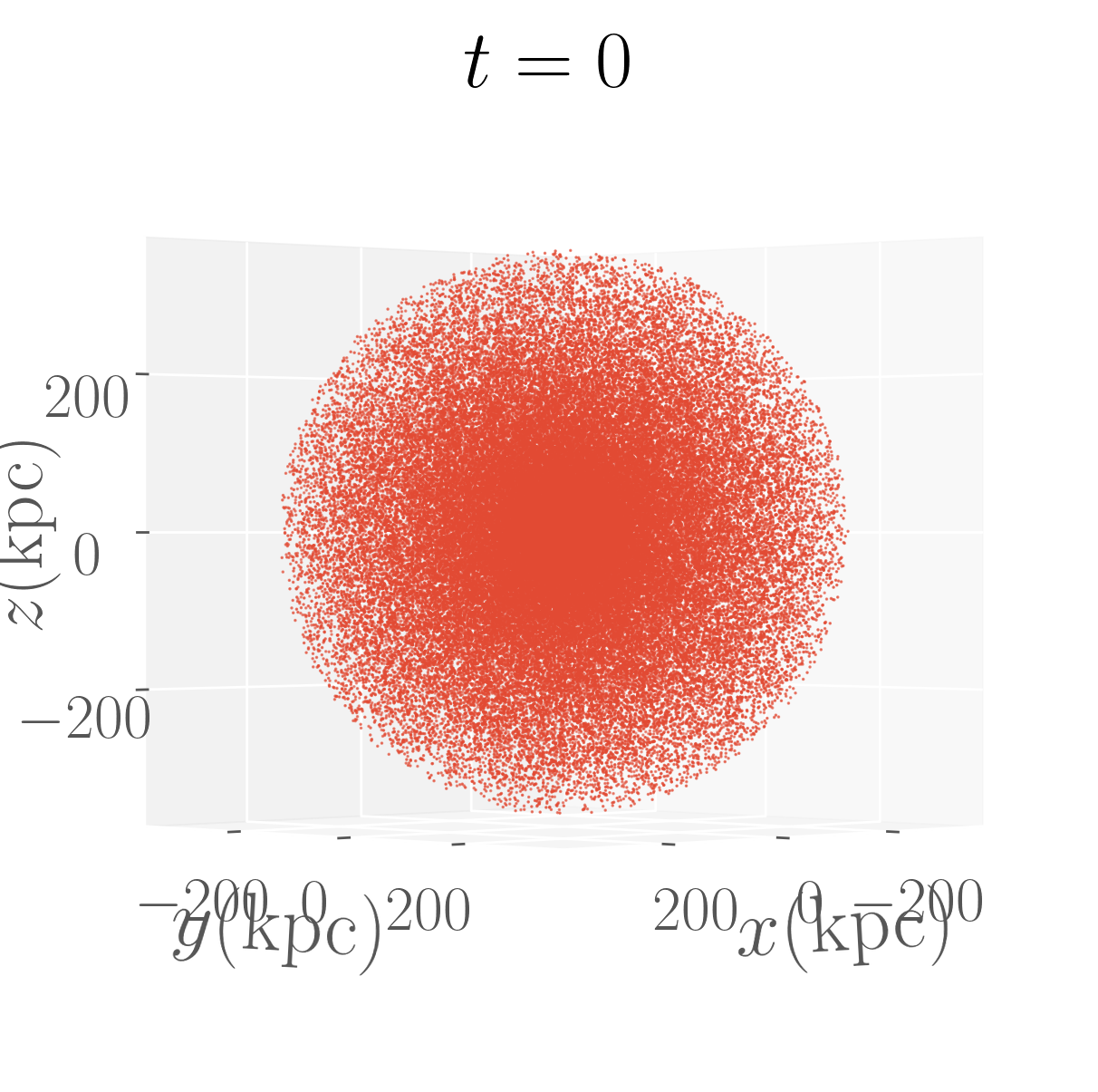}
\includegraphics[width=0.21\textwidth]{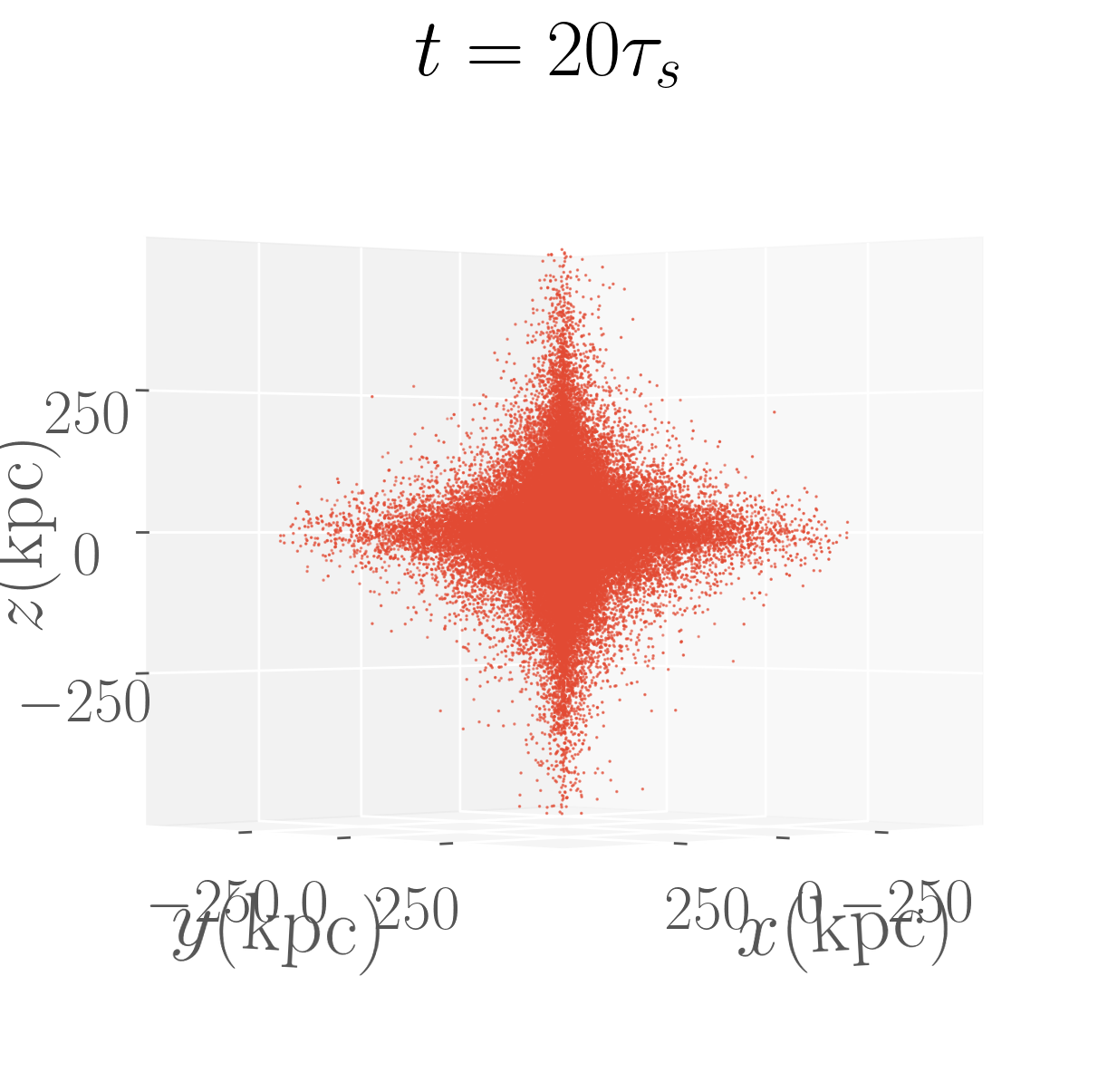}
\includegraphics[width=0.21\textwidth]{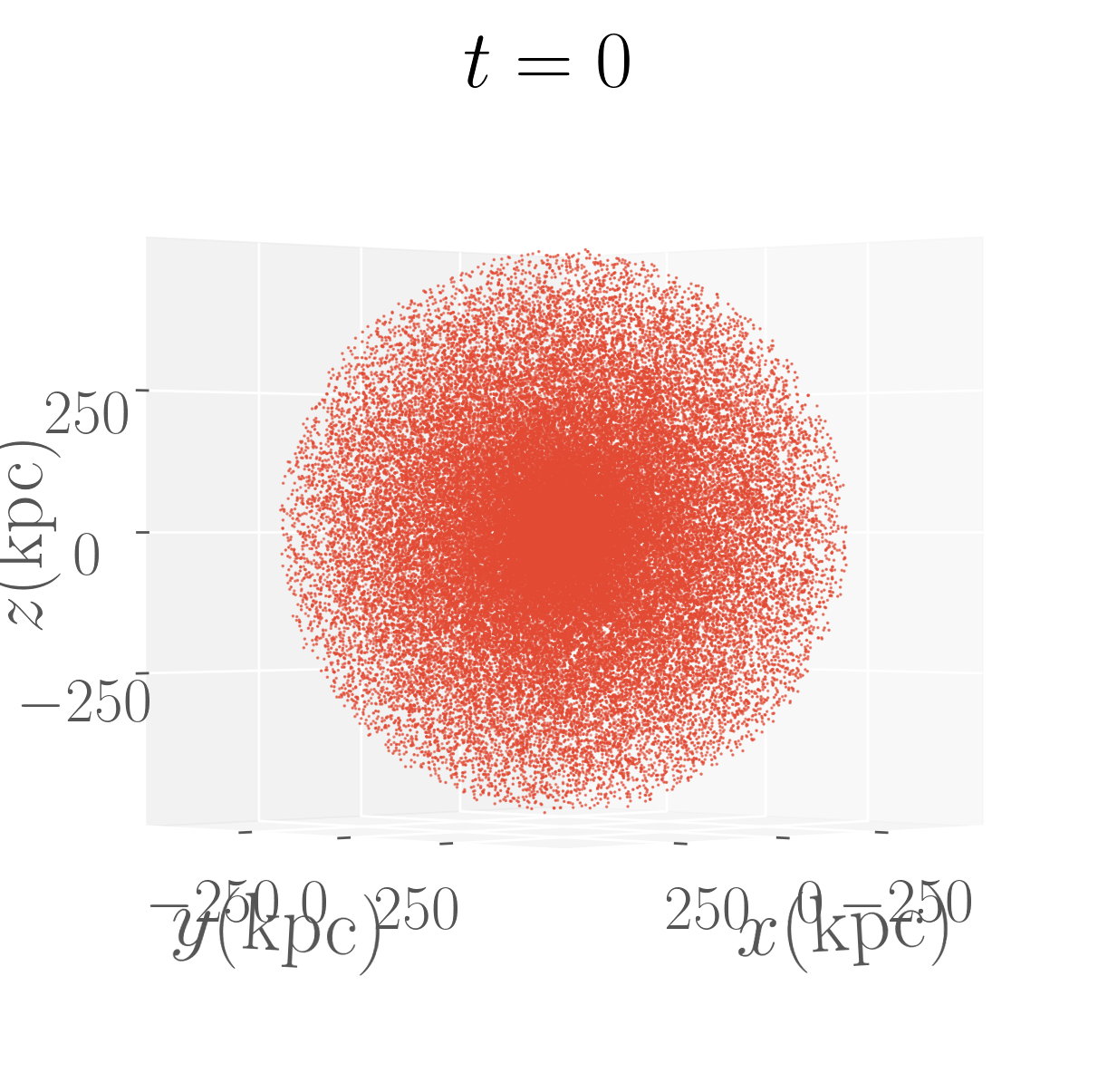}
\includegraphics[width=0.22\textwidth]{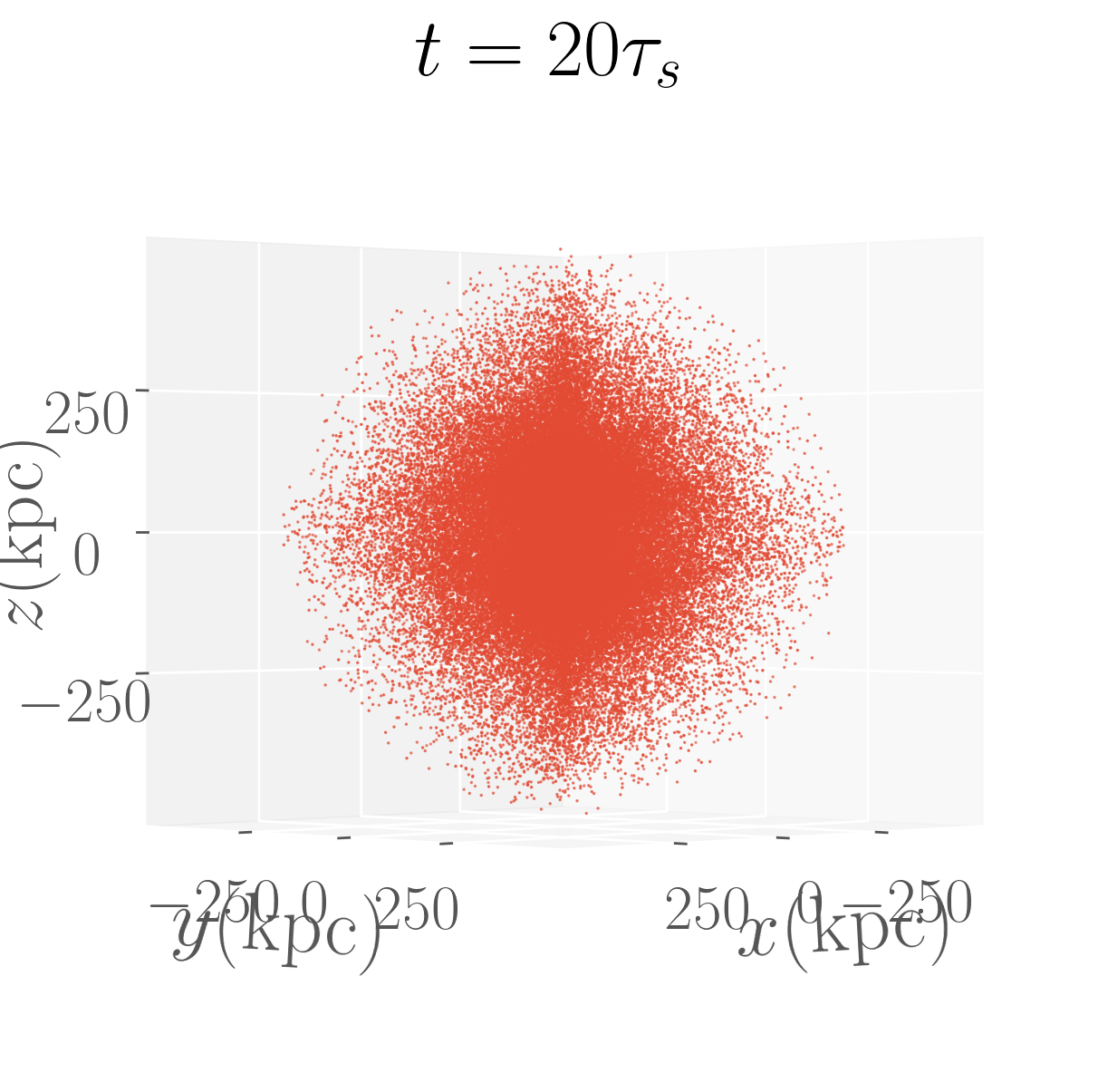}
\caption{Spatial distribution of $10^5$ test particles at initial time, and after $20\tau_s$. At the top we show the position of test particles for the monopole dominated configuration with $M_{100}/M_{210} = 3$.
At the bottom we show the position particles for the dipolar dominated case $M_{100}/M_{210} = 0.36$.}
\label{fig:random}
\end{figure}

The particles travel on the gravitational potential sourced by the multistate halo in a wide variety of trajectories. { Given the randomness of the initial conditions, the regions where the particles accumulate the most, will also be those regions where a single particle has the bigger likelihood to reside.  We define the evolving timescale $\tau_s$ of the system, as the time it takes  a test particle, initially located on the equatorial plane at a distance $\lambda \hat{\mu}r=4$ from the origin, on a circular trajectory, with an initial velocity equal to a quarter its escape velocity, to complete an orbit. In physical units, this timescale takes the value of 1.8 and 47Gys, for the monopole and dipole dominated configurations respectively.

In Fig.~\ref{fig:random} we show the spatial distribution of particles at initial time and after evolving during a sufficiently long time of 20$\tau_s$, for the monopole and dipole dominating  configurations. The particles distribute anisotropically and concentrate mainly around the equatorial plane of the configuration and along the $z-$axis, an effect produced by the density blobs associated to the dipolar contribution to the density.}

It can be seen that in the monopole-dominated configuration the particles distribute in a star-like shape at large radii, but most of them remain concentrated around the center within a sphere of radius $\lesssim 200 \mathrm{kpc}$. In contrast, for the dipole-dominated configuration the particles seem to be distributed more symmetrically around the dipole axis, although still retaining a spherical shape at radii $\lesssim 200 \mathrm{kpc}$.

\begin{figure*}
\centering
\includegraphics[width=0.4\textwidth]{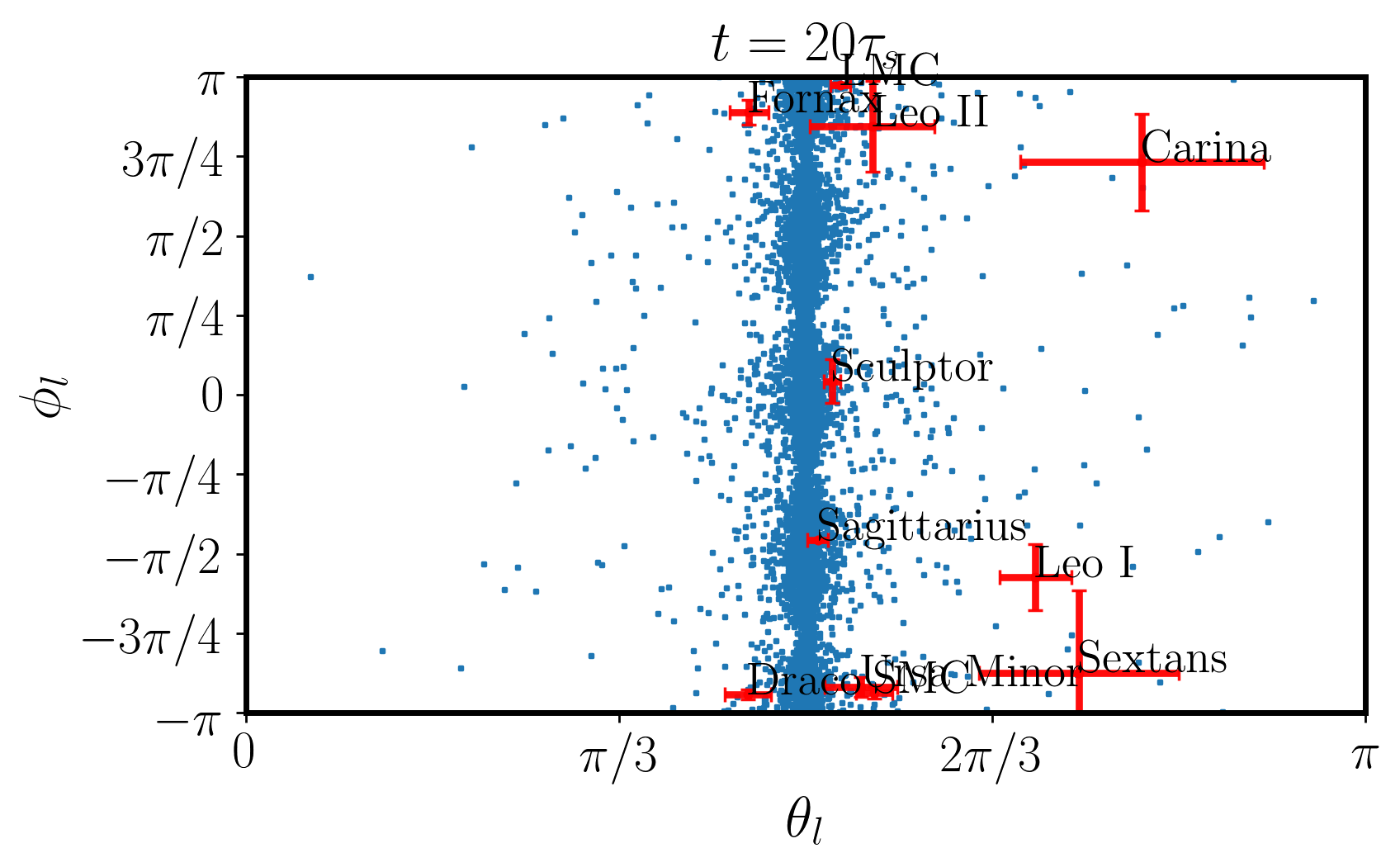}
\includegraphics[width=0.5\textwidth]{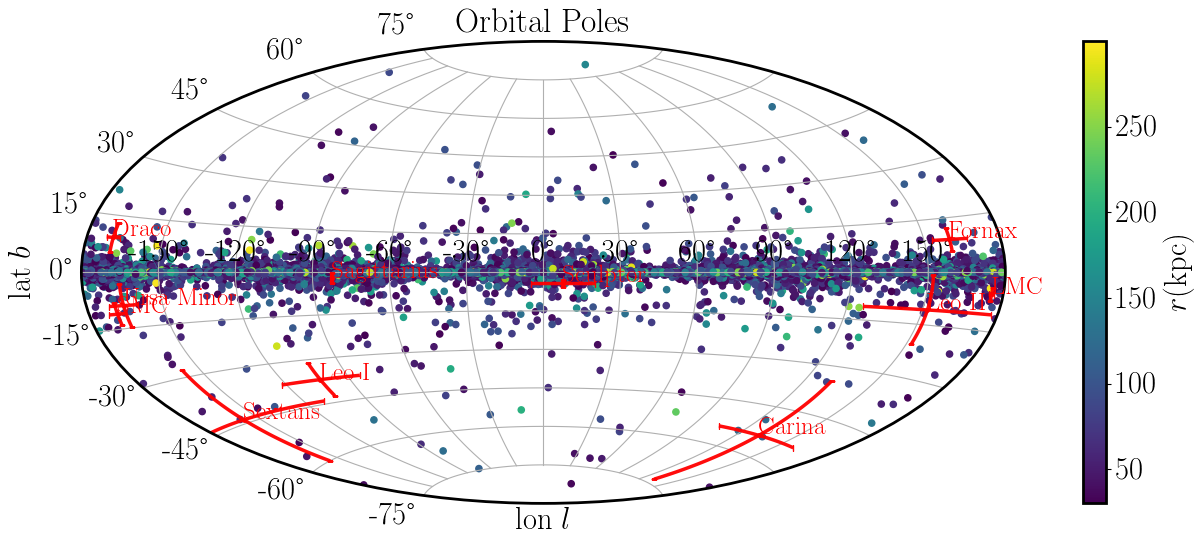}
\includegraphics[width=0.4\textwidth]{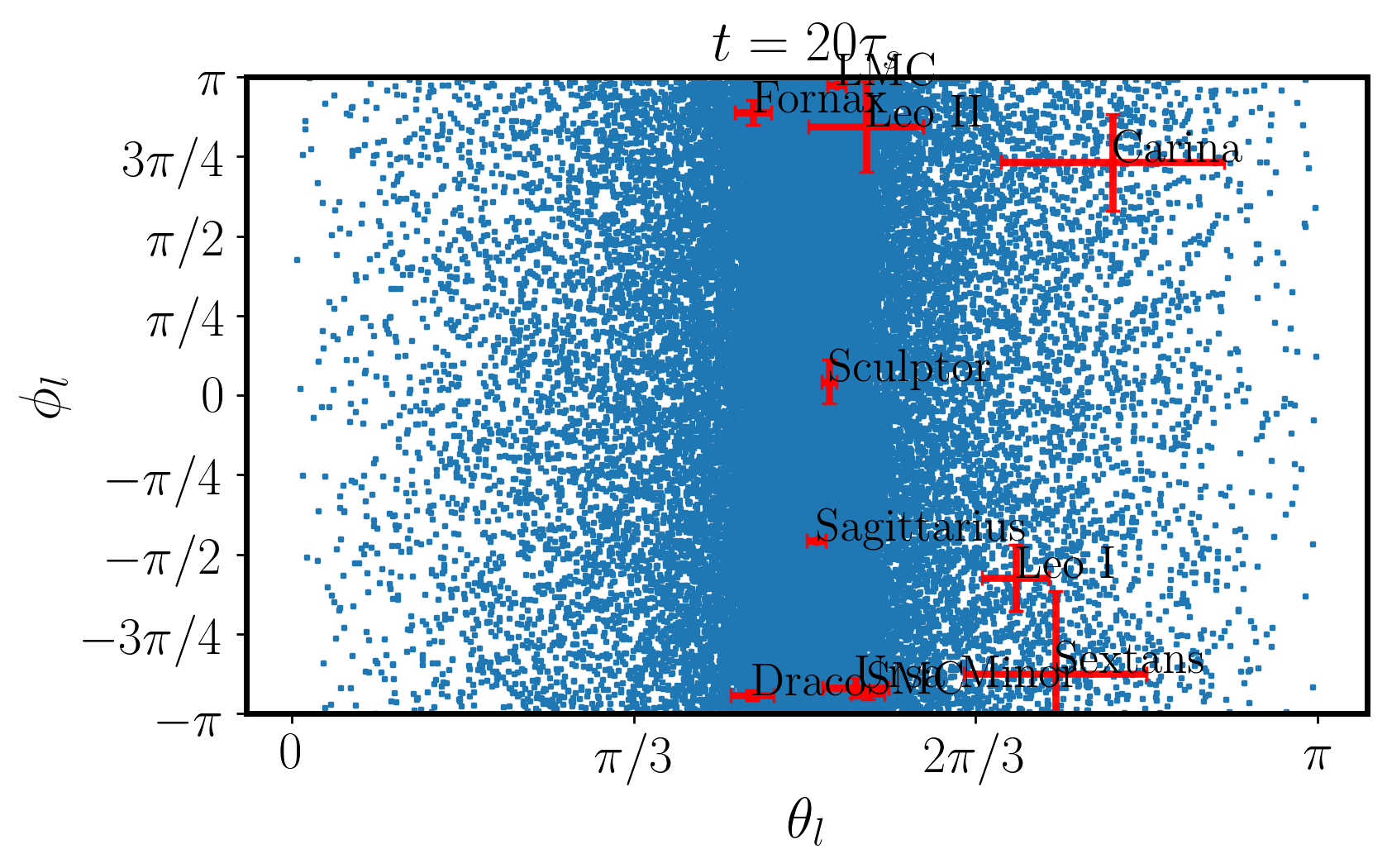}
\includegraphics[width=0.5\textwidth]{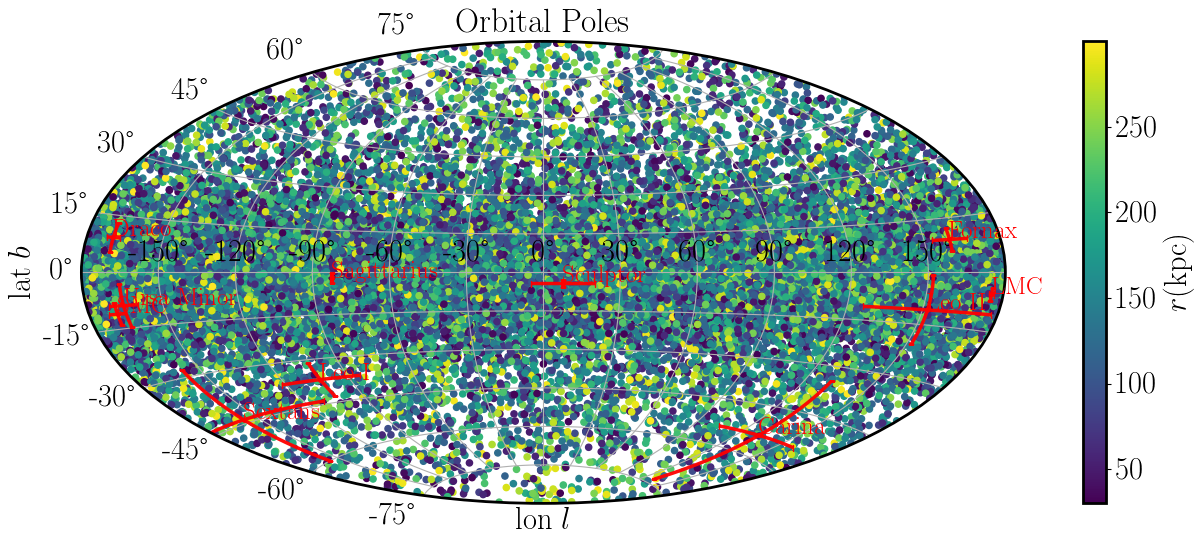}
\caption{{ Polar angle and azimuthal angle (left panel) and longitude and latitude (right panel) in galactocentric coordinates of the orbital poles (angular momentum) of the test particles that after $t = 20 \tau_s$ are in a distance range $r\in (30,300)$kpc, for the monopole dominated (upper row) and dipole dominated (bottom row) configurations. The red markers are the orbital poles of the Milky Way classical satellites.}}
\label{fig:orbital_poles}
\end{figure*}

For a comparison with observations of the Milky Way satellite galaxies, we show in Fig.~\ref{fig:orbital_poles} the orbital poles (angular momentum per unit mass) of the test particles in both monopole and dipole-dominated configurations at  $t = 20\tau_s$ in the range $r\in (30,300) \, \mathrm{kpc}$ from the origin, and the orbital poles of the Milky Way classical satellites, as calculated from the data reported in~\cite{Pawlowski:2019bar}. In both cases the { orbital poles of the} particles seem to distribute around a vertical column, being wider the one from the dipole-dominated configuration, which would in turn be more compatible with the data points of the satellite galaxies. In the right panels we show the same distributions but in terms of galactocentric longitude and latitude. Again, 
{ the orbital poles of the particles} remain distributed in a disc, but the one from the dipole dominated configuration is thicker. 

\begin{figure*}
 \centering
\includegraphics[width=0.19\textwidth]{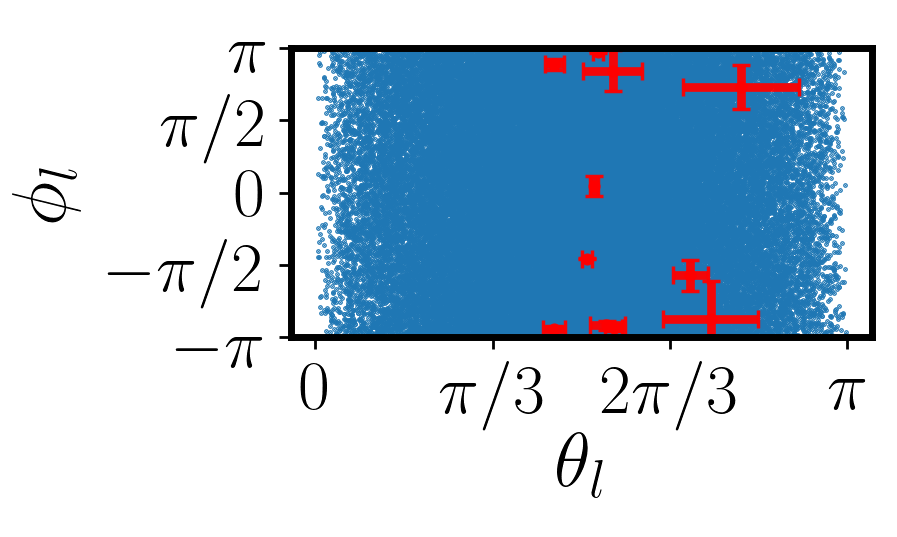}
\includegraphics[width=0.19\textwidth]{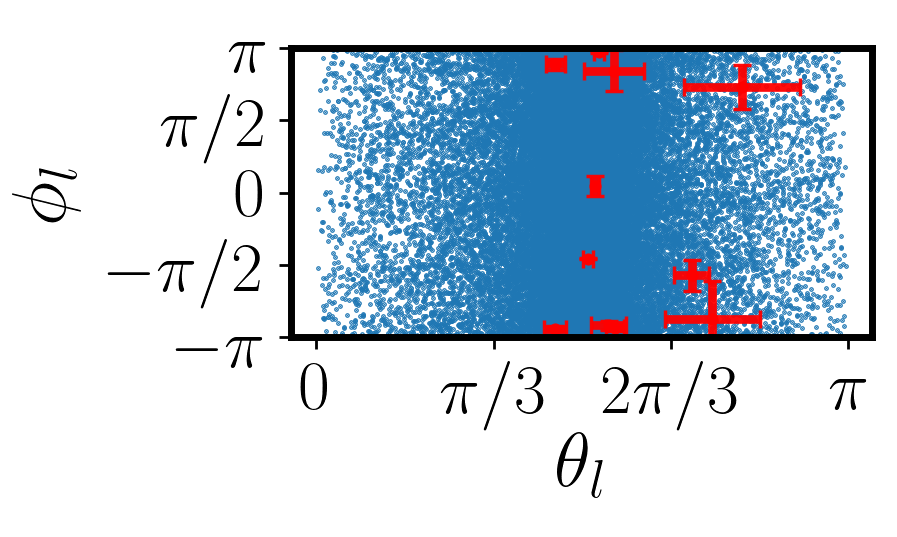}
\includegraphics[width=0.19\textwidth]{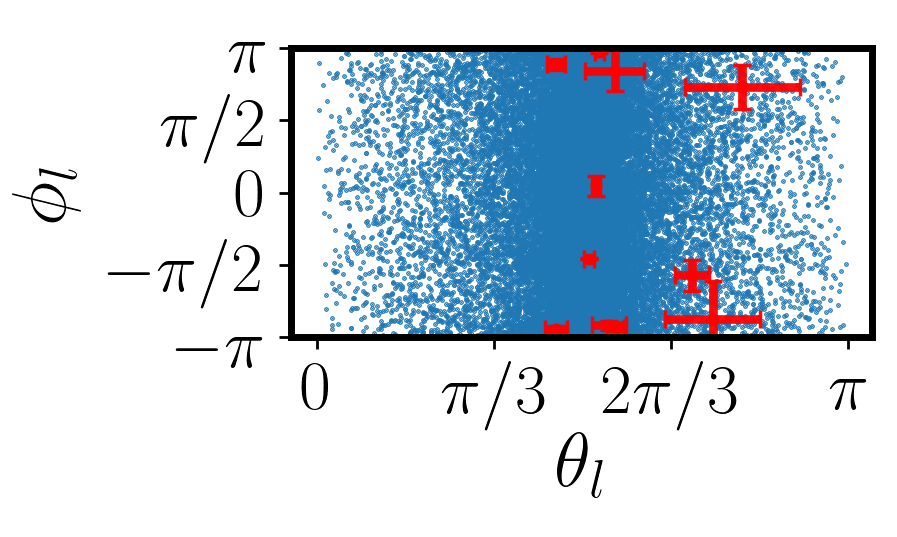}
\includegraphics[width=0.19\textwidth]{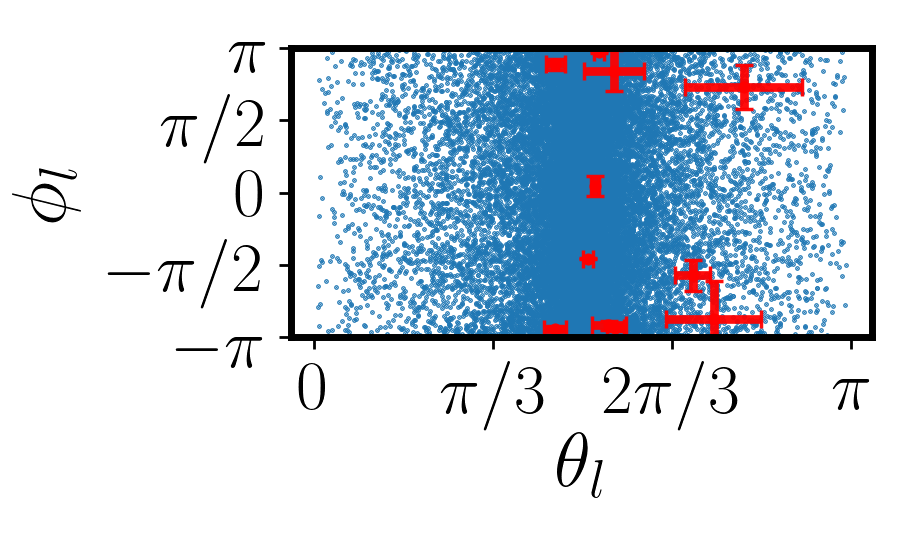}
\includegraphics[width=0.19\textwidth]{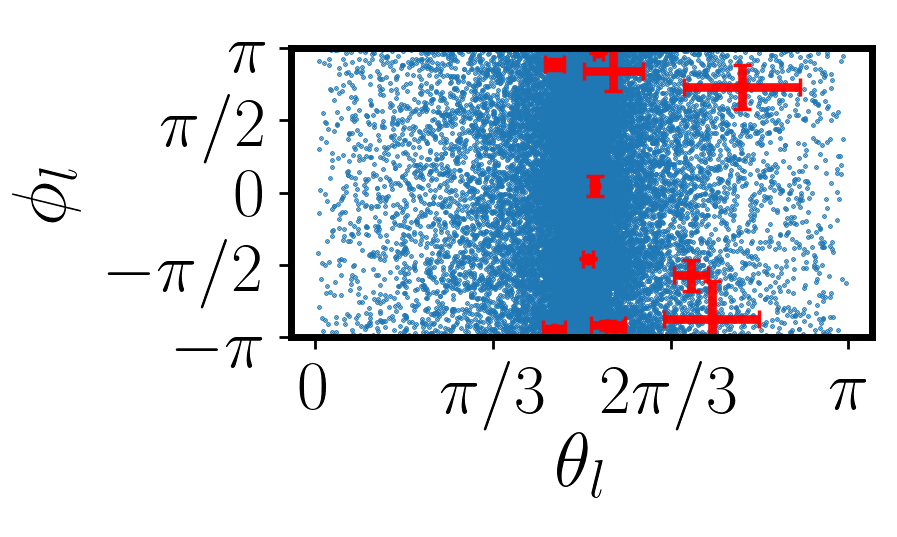}
\includegraphics[width=0.19\textwidth]{images/Firsttau/lth_lph_6_t0.png}
\includegraphics[width=0.19\textwidth]{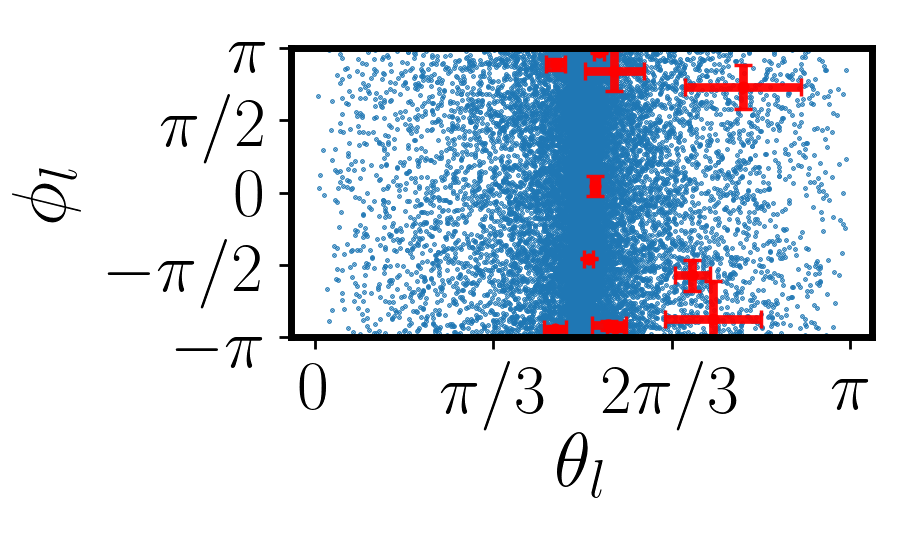}
\includegraphics[width=0.19\textwidth]{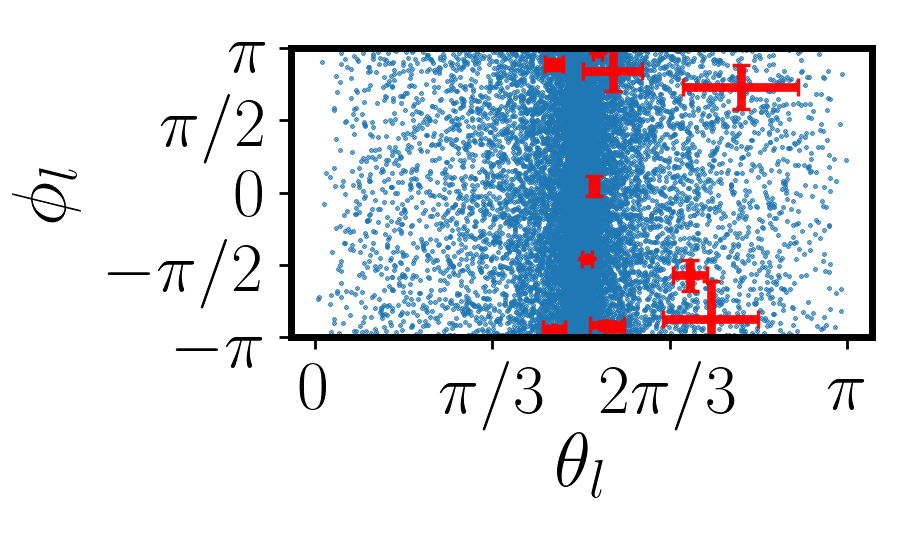}
\includegraphics[width=0.19\textwidth]{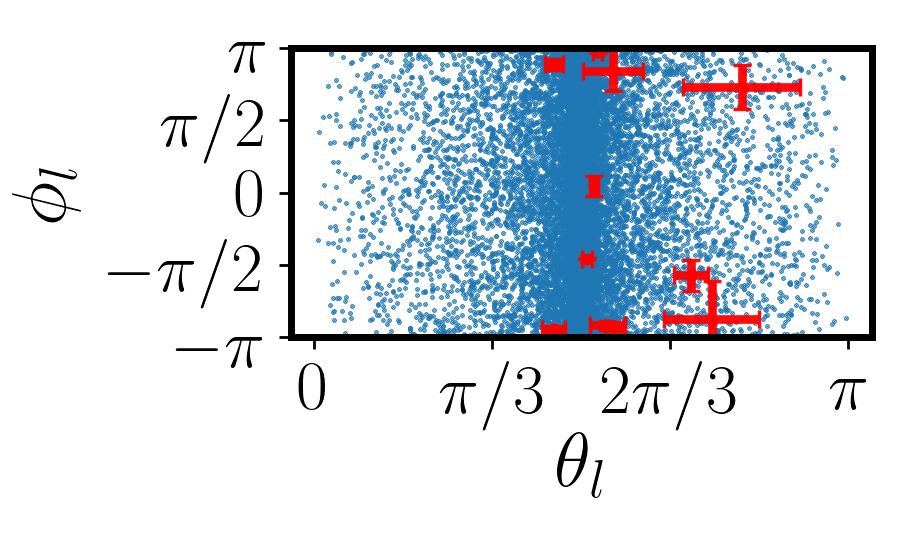}
\includegraphics[width=0.19\textwidth]{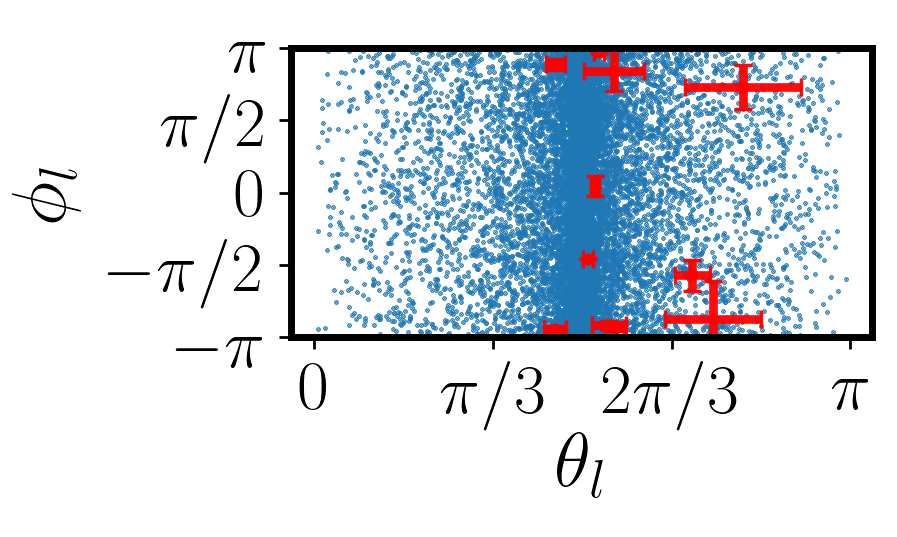}
 \caption{Snapshots of the angular poles for all the test particles under the potential of the monopole dominated configuration. The first row shows snapshots every 0.25$\tau_s$. The second row shows pictures every $4\tau_s$. Red markers are the orbital poles of the Milky Way classical satellites. In the monopole dominated configuration $\tau_s\simeq 1.8$Gyr, which implies that after 7.2 Gyrs the orbital poles distribution becomes nearly stationary.}
 \label{fig:ang_poles_snapshots}
 \end{figure*}

In Fig.~\ref{fig:ang_poles_snapshots} we show the evolution of the orbital poles of all the test particles for the monopole dominated configuration. The first 5 panels show snapshots every 0.25$\tau_s$ until $t=\tau_s$. The bottom panels show  the evolution every $4\tau_s$. {Two properties are to be noticed, first that the orbital poles  accumulate around $\pi/2$ with the pass of time, and second, the distribution of orbital poles tends to be stationary after 4$\tau_s$, which indicates late-time attractor properties.} 

\section{Consistency checks \label{sec:consistency}} 

{ In order to verify whether the multipolar distribution is consistent with galactic rotation curves, we use a simple model of a galaxy, consisting of stellar disc, bulge and dark matter halo. The circular velocity of a particle due to these components is

\begin{equation}
    v(r) = \sqrt{v_h^2 + v_d^2 +v_b^2}
    \label{eq:circ_vel}
\end{equation}

\noindent where the subscripts $(h, d, b)$ stand for halo, disc and bulge respectively. In the Appendix \ref{appendixA} we indicate the specifics of the model used for each matter component.
}

\begin{figure}
\centering
\includegraphics[width= 8cm]{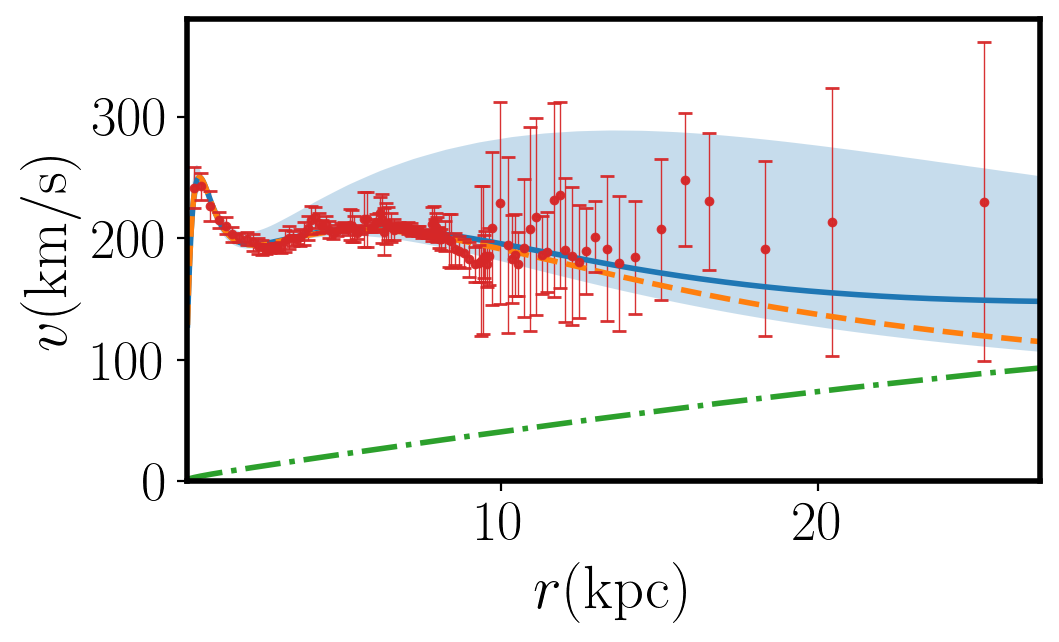}
\includegraphics[width= 8cm]{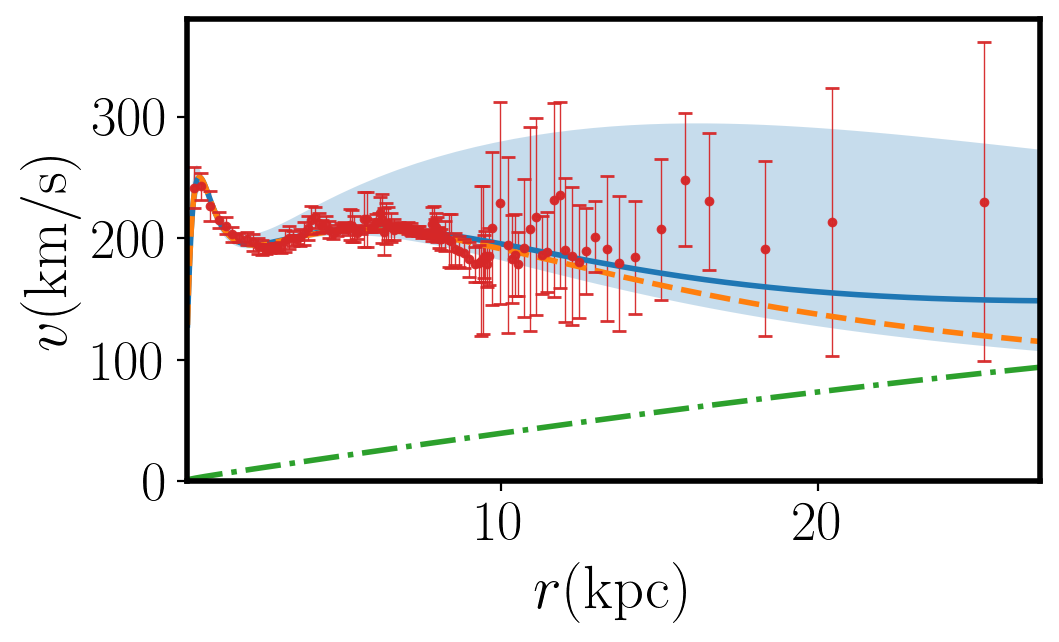}
\caption{{ Fit of the rotation curve of the Milky Way (blue solid line). The disc and bulge are modeled with exponential profiles (orange dashed line), and the dark matter halo (green dash-dotted line) with the multi-state configuration density $|\Psi_{100}|^2 + |\Psi_{210 }|^2$.
 Data points are taken from \cite{sofue2012}. On the top we show the monopole dominating configuration $M_{100}/M_{210} = 0.36$ and in the bottom the dipole dominating configuration $M_{100}/M_{210} = 3$.}}
\label{fig:Rotation}
\end{figure}

{ Using the Markov Chain Monte Carlo (MCMC) method, we fit the measurements of circular velocities of the Milky Way from \cite{sofue2012} by sampling the parameter space from uniform priors. We use $10^4$ steps with $30$ \% burn in and 300 walkers to sample the parameter space. The results for each one of the varied parameters were calculated using the LMFIT \citep{lmfit} and EMCEE \citep{emcee} PYTHON packages, and their values are shown in Table \ref{tab:resfits}. The output values from the MCMC method are consistent with other studies of the Milky Way with different DM models. In particular, the values of the SFDM parameters, the boson mass $\mu$ and the scaling parameter $\lambda$, are the same used in our studies of particle trajectories in Sec.~\ref{sec:particles}}. 

In Fig. \ref{fig:Rotation} we show the fit of the total rotation curve along with the contribution of the disc and bulge. On the top panel for the monopole dominating configuration and in the bottom one for the dipole dominating configuration. This shows that the multi-state configurations that have been studied allow the fitting of galaxy rotation curves.{ Also in Appendix \ref{appendixA} we show the posterior distribution of the fitting parameters.}

\begin{table}[]
\caption{Fit results of the parameters. Columns 3-5: mean, one and two $\sigma$ spread of the mean for the dipole dominating configuration respectively. Columns 6-8 same as in 3-5 but for the monopole dominating configuration.}
\label{tab:resfits}
\begin{tabular}{|cc|lll|lll|}
\hline
\multicolumn{1}{|l}{} & \multicolumn{1}{l|}{} & \multicolumn{3}{c|}{Dipole dominated}                                                     & \multicolumn{3}{c|}{Monopole dominated}                                                   \\ \cline{3-8} 
Name                  & Units                 & \multicolumn{1}{c}{Mean} & \multicolumn{1}{c}{$1\sigma$} & \multicolumn{1}{c|}{$2\sigma$} & \multicolumn{1}{c}{Mean} & \multicolumn{1}{c}{$1\sigma$} & \multicolumn{1}{c|}{$2\sigma$} \\
(1)                   & (2)                   & \multicolumn{1}{c}{(3)}  & \multicolumn{1}{c}{(4)}       & \multicolumn{1}{c|}{(5)}       & \multicolumn{1}{c}{(6)}  & \multicolumn{1}{c}{(7)}       & \multicolumn{1}{c|}{(8)}       \\ \hline
$\lambda$             & $10^{-3}$             & 0.94                     & 0.3                           & 0.64                           & 0.59                     & 0.19                          & 0.39                           \\
$\hat{\mu}$           & 1/kpc                 & 70.733                   & 44.141                        & 68.744                         & 59.868                   & 39.485                        & 67.336                         \\
$\mu c^2$             & $10^{-25}$ eV         & 4.5228                   & 2.8225                        & 4.3957                         & 3.8281                   & 2.5248                        & 4.3056                         \\
$M_d$                 & $10^{10}M_\odot$      & 6.7594                   & 0.3385                        & 0.7177                         & 6.7494                   & 0.3484                        & 0.7438                         \\
$a_d$                 & kpc                   & 3.1269                   & 0.1199                        & 0.2491                         & 3.1265                   & 0.1207                        & 0.2521                         \\
$M_b$                 & $10^{10}M_\odot$      & 0.9737                   & 0.0454                        & 0.0909                         & 0.9733                   & 0.0453                        & 0.0904                         \\
$a_b$                 & kpc                   & 0.1353                   & 0.0118                        & 0.0237                         & 0.1354                   & 0.0119                        & 0.0238                         \\ \hline
\end{tabular}
\end{table}

{ Another important check consists in showing that a disc like structure, perhaps that of a disc galaxy, is not destroyed by the multistate configuration. For this, {as an approximation, neglecting the interaction between particles as occurs in the case of stellar discs}, we follow the trajectory of $10^4$ particles initially distributed in a double exponential disc with extension $\lambda \hat{\mu} R = 1$. Initially the particles have only circular velocity, and during the evolution the potential due to the multistate configuration influences the trajectories.}

\begin{figure}[tp!]
\centering
\includegraphics[width=0.23\textwidth]{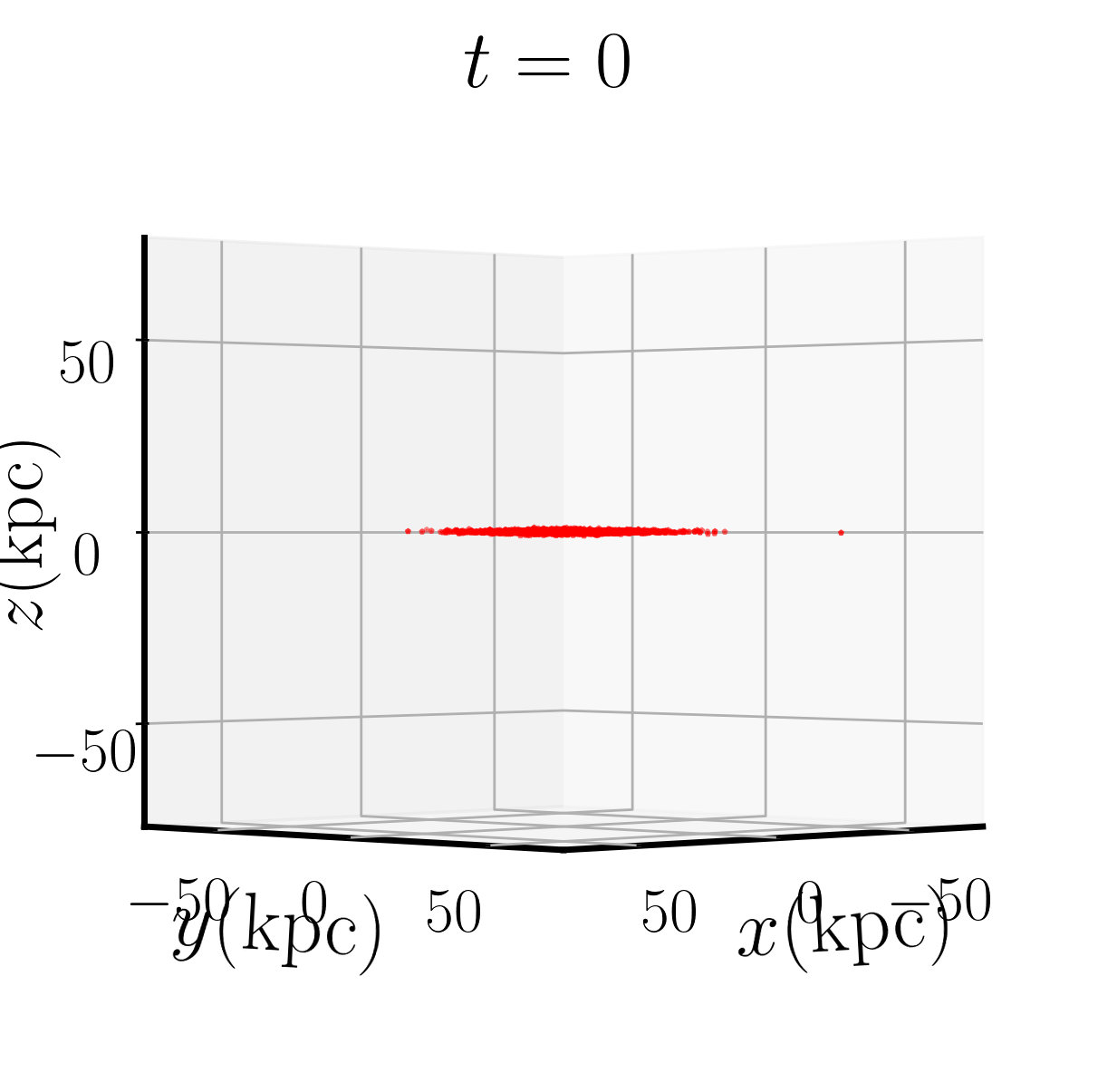}
\includegraphics[width=0.23\textwidth]{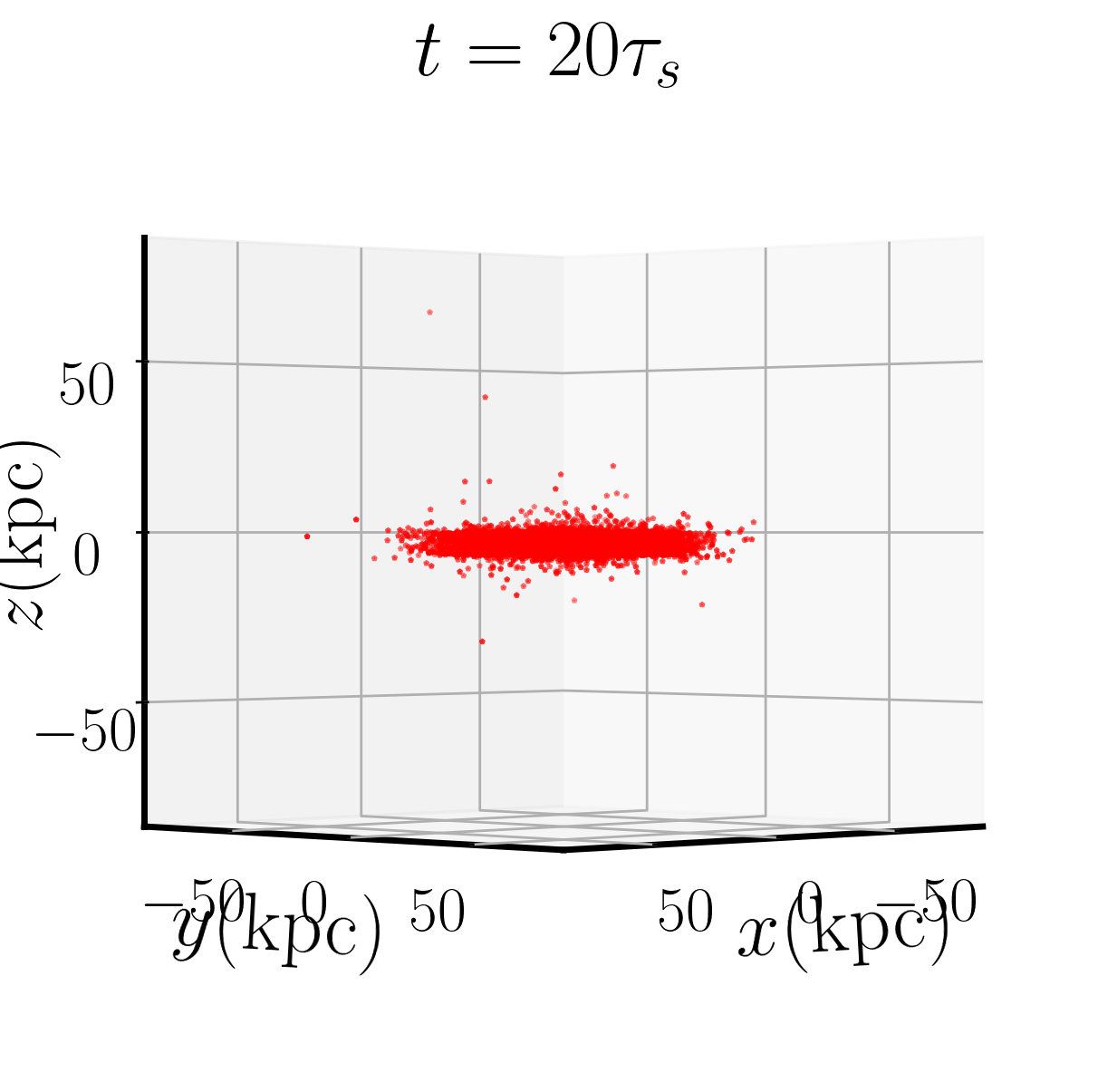}
\includegraphics[width=0.23\textwidth]{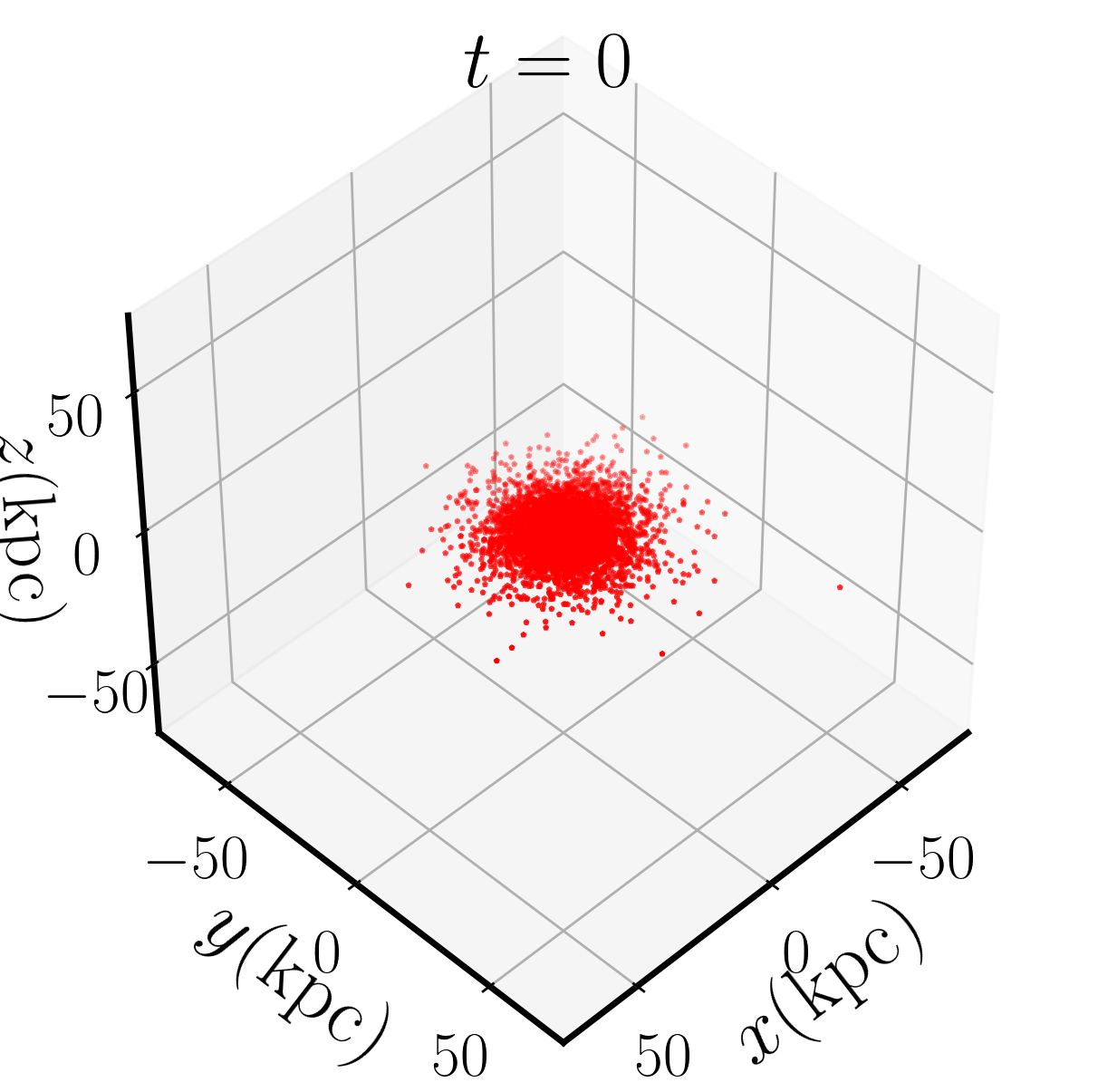}
\includegraphics[width=0.23\textwidth]{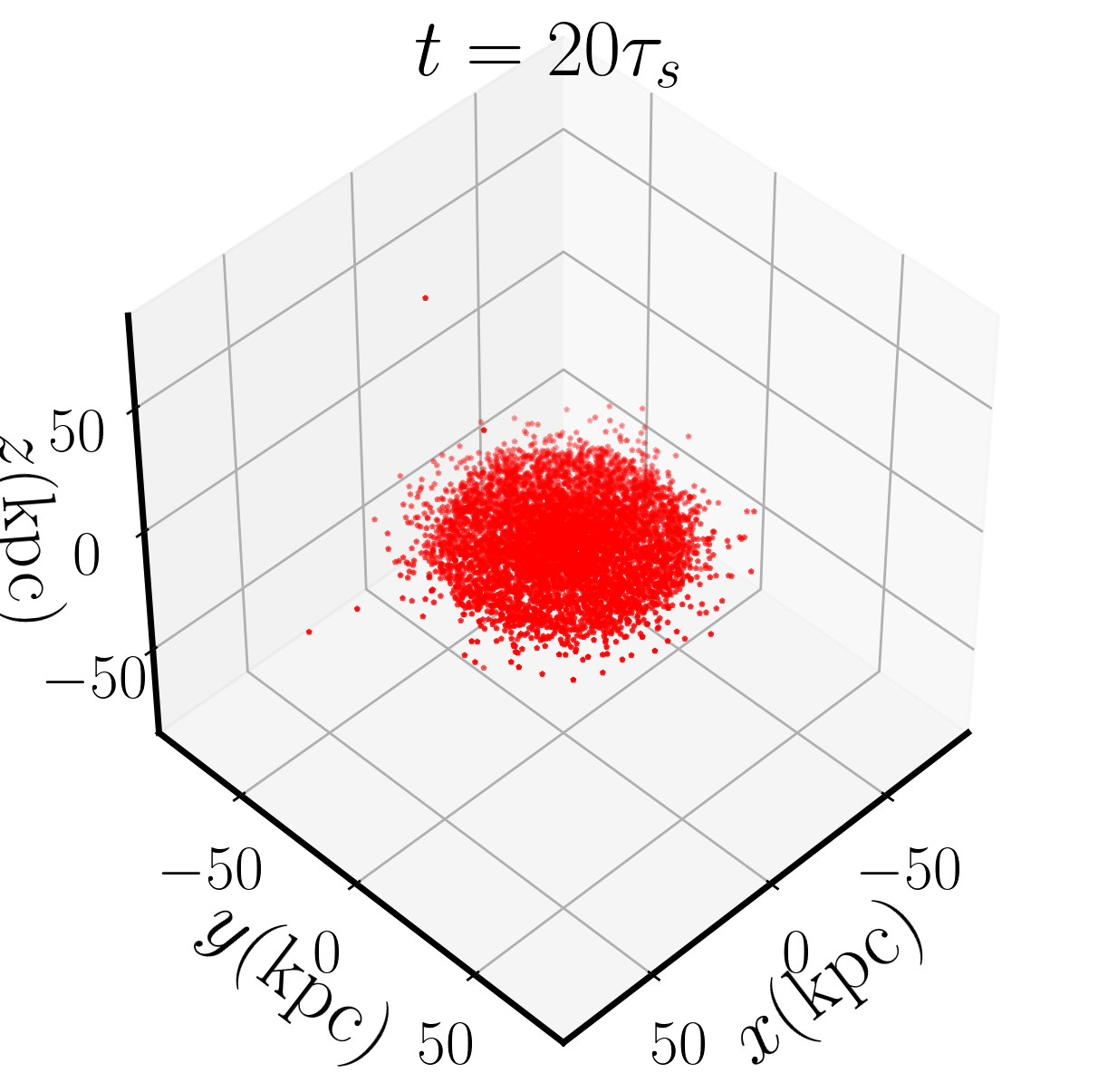}
\caption{ At the left we show the distribution of particles in a double exponential disc distribution at the left at initial time with two perspectives. The initial radius of the disc is $\sim 30$kpc. At the right we show the evolved particles after $t =20 \tau_s$, showing an expanded disc of radius $\sim 50$kpc.}
\label{fig:random4}
\end{figure}

We show the distribution of particles after $t = 20 \tau_s$ {for the monopole-dominated case,} that keeps the disk-shaped distribution as shown in Fig. \ref{fig:random4}, which is thicker due to the attraction produced by the blobs of the dipolar component, but the disc is  not destroyed. {For the dipolar-dominated configuration, the influence of the dipolar component destroys the disc.} We also performed tests with tilted discs and we found they are actually destroyed by the axial position of the potential minima of the dipolar contribution. This limitation is important for the interpretation of our results.

The configuration used here makes test particles to concentrate in a nonisotropical fashion, mainly near the equatorial plane and the poles. Nevertheless, the viability of multipolar configurations as a halo model, depends on whether  they are long-lived, being the minimal condition for them to be stable. In order to check this condition, following the recipe in \cite{Nuevo} we evolved the configurations by solving the fully time-dependent equations~\eqref{eq:gpp} using a multistate generalization of the code that solves the GPP system \cite{Nkode3d}. It was found that these configurations oscillate around a virialized state and are long-lived. The strongest check was that the oscillation frequency of the wave functions coincides with those found when solving the eigenvalue-problem from Eqs.~\eqref{eq:gpp} and~\eqref{eq:stationarywfs}. 

\begin{figure}[tp!]
\centering
\includegraphics[width=8cm]{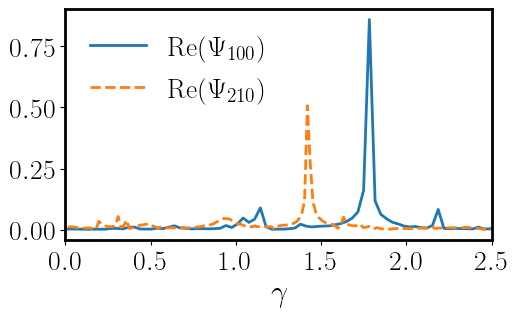}
\includegraphics[width=8cm]{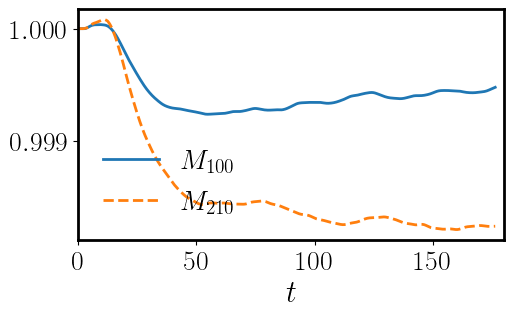}
\caption{
Top: Fourier transform of the maximum of the real part of the ground state $\Psi_{100}$ and excited state $\Psi_{210}$. The peaks appear near the eigenfrequencies $\gamma_{100}=1.8$ and $\gamma_{210}=1.42$, which confirms the wave functions oscillate with the eigenfrequency throughout the evolution. Bottom: we show $M_{100}$ and $M_{210}$ as function of time, normalized to their initial values, and showing that their values remain nearly constant within less than 0.2
\%.
}
\label{fig:evolution}
\end{figure}

We ran the dipole-dominating configuration used in our analysis during a time window of 200 periods of the wave functions $\Psi_{100}$. In Fig. \ref{fig:evolution} we show the Fourier transform of the maximum of the individual wave functions $\Psi_{100}$ and $\Psi_{210}$.  The peak frequencies are consistent with the frequencies found by solving for the stationary configuration $\gamma_{100}=1.8$ and $\gamma_{210}=1.42$. We also show that $M_{100}=\int |\Psi_{100}|^2d^3 x$ and $M_{210}=\int |\Psi_{210}|^2d^3 x$ change only less than 0.2\%, indicating that the evolution is nearly unitary. These results show that the configuration is long lived.

{ In Appendix \ref{appendixB} we present a set of other consistency checks more related to the methods used to analyze the dynamics of test particles.}


\section{Discussion}

{ Our results indicate that multi-state, (1,0,0)+(2,1,0) solutions of the GPP system, induce randomly initialized test particles to distribute anisotropically, with high concentrations at the poles and polar angle distribution concentrating toward the equatorial plane of the system, in the asymptotic time.}

{ The interpretation of this result is that a single particle, interpreted as a satellite galaxy hosted in a galaxy with a multi-state ultra-light bosonic dark matter halo, farther than 30kpc from the galactic origin, would be more likely to orbit with polar angle near $\pi/2$, on flat trajectories.}

{ We have shown how the particles distribute for two sample configurations with different density-mode domination, with a distinct polar angle distribution. Between these two examples there is a continuous universe of mass ratio $M_{100}/M_{210}$ whose effects may vary continuously, and thus potentially useful to study each particular case of host galaxy.}

{ Our analysis is supported by a set of consistency checks, including the viability of the multistate halo as a long-living self-gravitating structure, consistency of the multistate solutions with rotation curves and stability of disc distributions centered at the equator.}

{Now, not all the galaxies mentioned have satellites with polar angles near the equatorial plane of the host galaxy, Andromeda for one case. This, together with the fact that our multistate halos destroy tilted disc configurations of test particles, tilted with respect to the equatorial plane which is perpendicular to the axis of the dipole, make  the model to seem in contradiction with these observations. Nevertheless, our results are valid in the long-term, which means that eventually the polar angles of satellites in Andromeda should approach $\pi/2$ as time evolves.}

{ Finally, the motion of test
particles traveling on top of these multistate configurations is different from the results obtained if a nonspherical CDM halo is assumed instead.} For this we analyzed the distribution of particles and orbital poles moving on a NFW distorted halo as described in Appendix \ref{appendixNFW}. These differences are worth analyzing in a detailed parameter space exploration of both, the $M_{100}/M_{210}$ ratio and on the distortion parameters of a triaxial NFW halo.

{ That these multi-state configurations are equilibrium solutions of the GPP equations, indicates that the ultralight bosonic dark matter has potential to explain the VPOS observations in the known cases of the Milky Way, M31, CenA and other possible cases to come. A neat property is that multipolar solutions are natural to this model due to the properties of the GPP system of equations, which in turn results from the bosonic nature of the SFDM candidate.
}


\acknowledgments
J.S. acknowledge financial support from CONACyT doctoral fellowship.
This work was partially supported by CONACyT M\'exico under grants CB-2011 No. 166212, CB-2014-01 No. 240512, CB-2017 No. A1-S-17899, 304001, Project No. 269652;
I0101/131/07 C-234/07 of the Instituto
Avanzado de Cosmolog\'ia (IAC) collaboration (http://www.iac.edu.mx/). 
This research received support by Conacyt through the 
Fondo Sectorial de Investigaci\'on para la Educaci\'on, grant No. 240512 (TM) and No. 258726 (FSG).
F.S.G. acknowledges support from grant No. 4.9-CIC of the Science Research Program of Universidad Michoacana, and the use of the Big Mamma cluster at the IFM-UMSNH, where the numerical work was carried out.
L.A.U-L. was partially supported by Programa para el Desarrollo Profesional Docente; and Direcci\'on de Apoyo a la Investigaci\'on y al Posgrado, Universidad de Guanajuato. 
VHR acknowledges support from the YCAA Prize Postdoctoral Fellowship. 

\appendix

\section{GALACTIC MODEL}
\label{appendixA}

The stellar disc is modeled using a razor-thin exponential disc profile whose surface mass density written in cylindrical coordinates $(\rho,\phi,z)$ is given by

\begin{equation}
    \Sigma_d (\rho) = \Sigma_0 e^{-\rho/a_d},
\end{equation}

\noindent where $a_d$ is the disc scale length,  $\Sigma_0$ is the surface density at distance $\rho=0$ from the origin, it is related to the total mass of the disc $M_d$ as $M_d = 2\pi\Sigma_0 a_d^2$. The circular velocity due to this density profile is \citep{freeman1970ApJ}
\begin{equation}
    v_{d} = \sqrt{\frac{G M_d y^2}{2 a_d} \left( I_0\left(\frac{y}{2} \right) K_0\left(\frac{y}{2} \right) - I_1\left(\frac{y}{2} \right) K_1\left(\frac{y}{2} \right) \right)},
\end{equation}
where $I_n$ and $K_n$ are the modified Bessel functions of the first and second kind, respectively, and we have defined $y\equiv r/a_d$.

The galaxy bulge is modeled using an exponential density profile \citep{vancou1,vancou2} written in spherical coordinates $(r, \theta,\phi)$ is given by:
\begin{equation}
    \rho_b (r) = \rho_0 e^{-r/a_b},
\end{equation}

\noindent where $a_b$ is the bulge scale length and $\rho_0$ is the central density, related to the total mass of the bulge $M_b$ by $\rho_0 = M_b/(8 \pi a_d^3)$. The circular velocity due to this profile is
\begin{equation}
    v_b = \sqrt{\frac{G M_b}{r} \left( 1 - \left(1 + \frac{r}{a_b} + \frac{r^2}{2a_b^2} \right)e^{-r/a_b}\right) }.
\end{equation}

\begin{figure*}
\centering
 \includegraphics[width=0.7\textwidth]{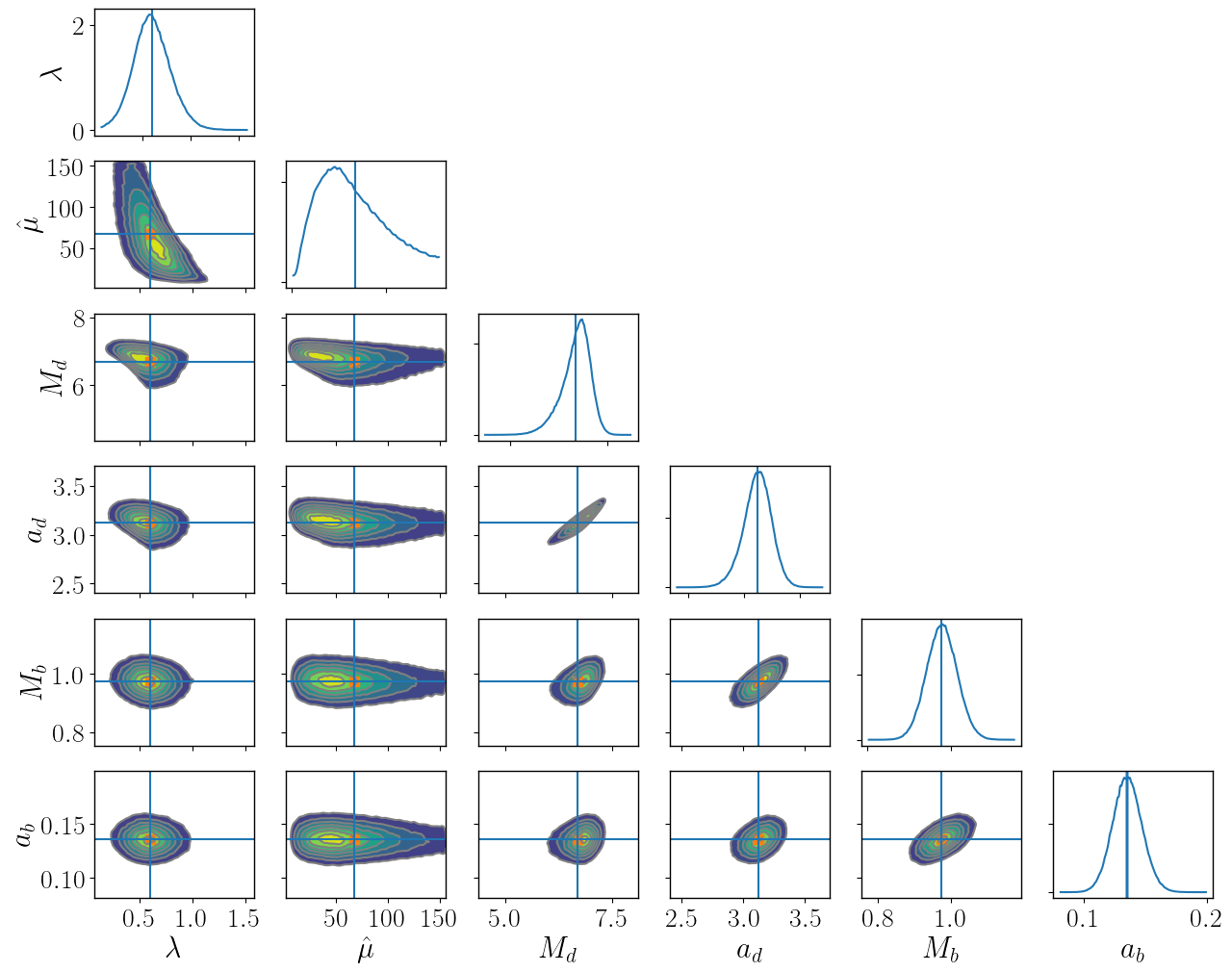}
 \includegraphics[width=0.7\textwidth]{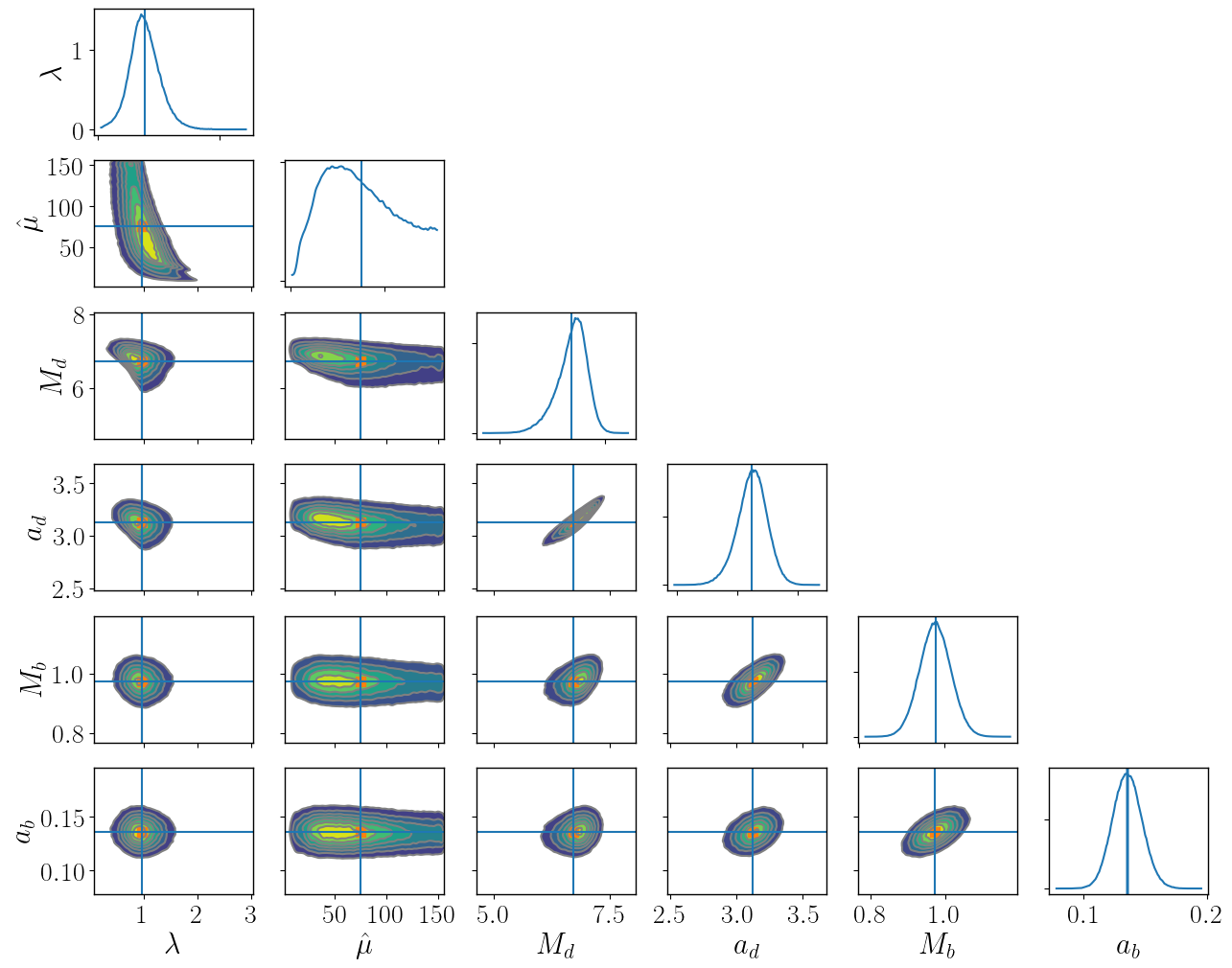}
\caption{Confidence intervals of the adjustment of parameters that best fit the Milky Way's rotation curve. In the top and bottom we show the plots for the monopole and dipole dominating cases respectively.}
\label{fig:posteriors}
\end{figure*}

We will use both of the multistate configurations we have been presenting, for these we have two parameters  to fit namely $\lambda$ and $\hat{\mu}$. The circular velocity of a particle due to these SFDM halos is given by

\begin{equation}
    v_h = c\lambda \sqrt{\rho \left. \frac{\partial V}{\partial \rho} \right\rvert_{z = 0}} , \label{eq:halo}
\end{equation}

\noindent which completes the information needed in Eq. (\ref{eq:circ_vel}). Finally, the  parameter estimates used to fit  the Milky Way rotation curve shown in Fig.~\ref{fig:Rotation}, have confidence intervals that are shown in Fig. \ref{fig:posteriors}.

\section{FURTHER DIAGNOSTICS}
\label{appendixB}

\begin{figure}[tp!]
 \centering
 \includegraphics[width=0.22\textwidth]{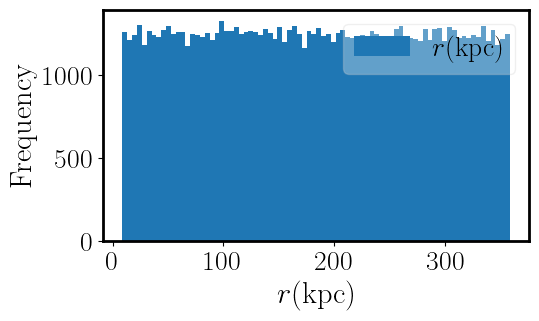}
  \includegraphics[width=0.22\textwidth]{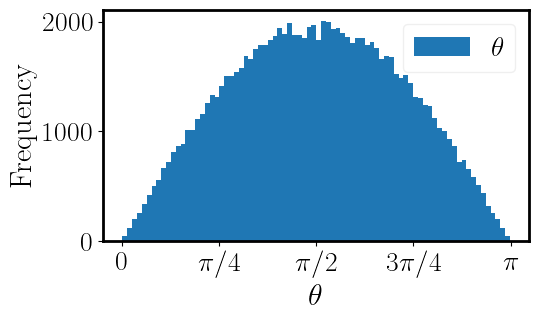}
  \includegraphics[width=0.22\textwidth]{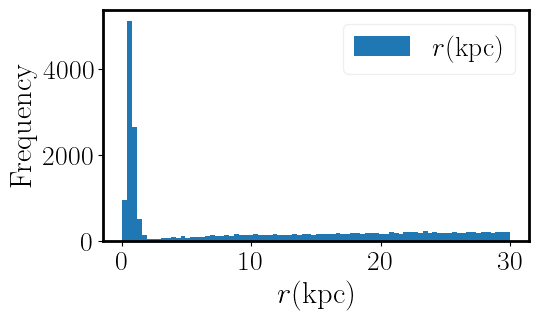}
  \includegraphics[width=0.22\textwidth]{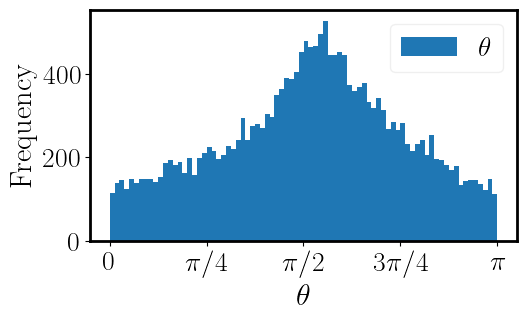} 
   \includegraphics[width=0.22\textwidth]{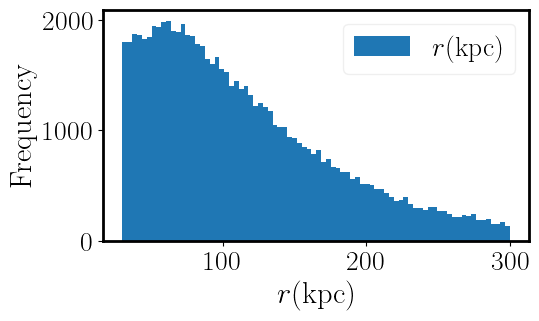}
   \includegraphics[width=0.22\textwidth]{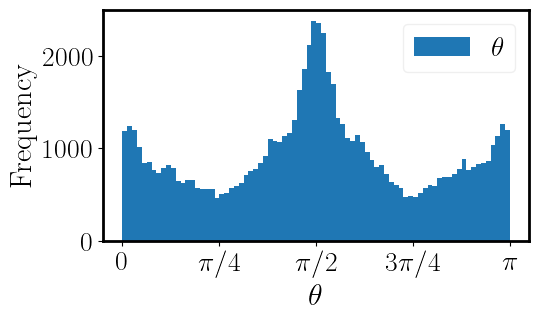} 
 \caption{ In physical units for $\mu = 10^{-25} \text{eV}/c^2$ we show the histogram of the $10^5$ test particles at initial time in the first row, and after $20\tau_s$ in second and third rows, for the monopole dominating configuration. The data is filtered by distances, in the second row we see the particles at short distances $r \in (0, 30)$kpc and in the third row the particles with at $r \in (30, 300)$kpc from the center of the configuration.}
 \label{fig:histopot6}
 \end{figure}

For a better understanding of the particle distribution, in Fig. \ref{fig:histopot6} we show the histogram of the radial distance $r$ of the particles and their polar angle $\theta$ after a lapse of 20$\tau_s$ for the monopole dominating configuration. The first row of plots corresponds to the distribution at initial time. In order to separate those particles within a distance of the order of the galaxy size from those at distances of order of distances corresponding to satellite galaxies, we show the histograms filtered by distances. In the second row we show the accumulation of test particles in the range $r \in (0, 30)$kpc and in the third row test particles in the range $r \in (30, 300)$kpc. In the later case we notice that particles distribute anisotropically at three preferential angles $\theta = 0, \pi/2$ and $\pi$. The interpretation of this result is that particles with random initial conditions will accumulate with bigger probability near these angles.

\begin{figure}[tp!]
 \centering
 \includegraphics[width=0.22\textwidth]{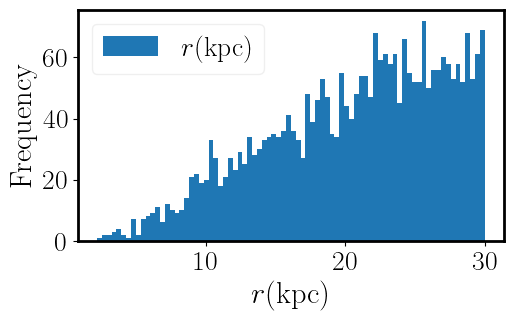}
 \includegraphics[width=0.22\textwidth]{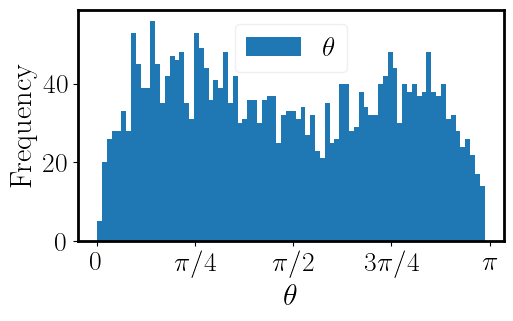}
 \includegraphics[width=0.22\textwidth]{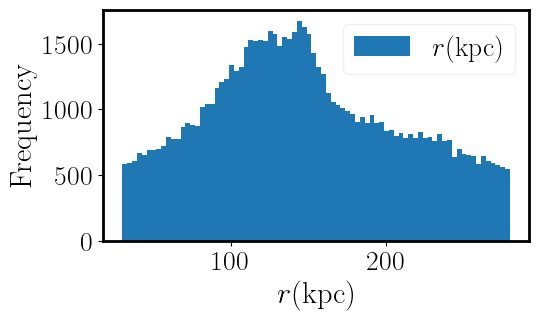}
 \includegraphics[width=0.22\textwidth]{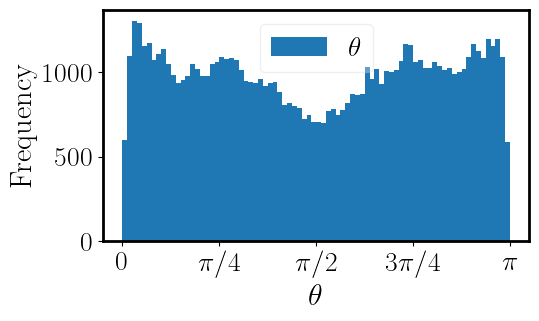}
 \caption{Histograms of accumulation of particles after 20$\tau_s$ on the radial and angular coordinates for the dipole dominating configuration. The anisotropy is not as clear as in the monopole-dominating case of Fig. \ref{fig:histopot6} as seen in the right panels. On the top we show histograms for particles within 30kpc and on the bottom within 30 and 280kpc.}
 \label{fig:histopot2}
 \end{figure}

In Fig.~\ref{fig:histopot2} we show the results for the dipole-dominating configuration. There are important differences, starting with the fact that particles do not accumulate near the origin, instead show a peak concentration around 30kpc due to the influence of the dipolar contribution, which is dominating. The concentration of particles in angles is also different, the distribution is not isotropic either, but the peaks are not as sharp as in the monopole-dominating scenario. This is an indication that the model can have restrictions to clearly explain the accumulation of particles at planes that pass near the poles.

\begin{figure}[tp!]
\centering
\includegraphics[width=6cm]{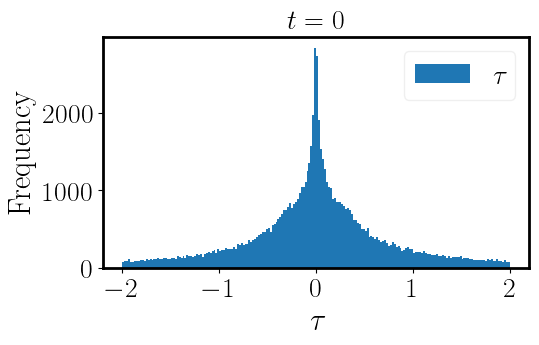}
\includegraphics[width=5.5cm]{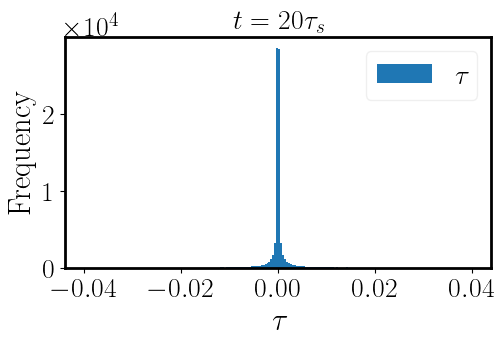}
\includegraphics[width=6cm]{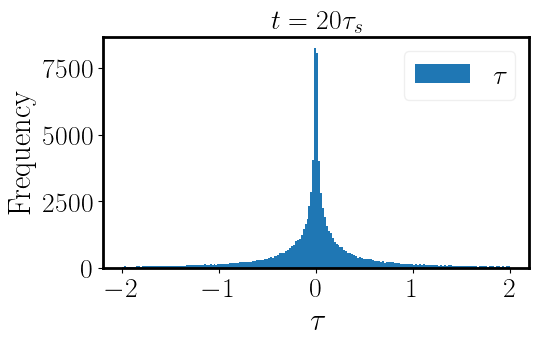}
\caption{Histogram of the torsion $\tau$ of trajectories of test particles. At the top we show the distribution of values at initial time. In the middle and bottom we show the result after $t=20\tau_s$ for the of the monopole and dipole dominated configurations respectively. }
\label{fig:torsion_core_domin}
\end{figure}

{\it Plane trajectories.} In our probabilistic approach, we have seen so far that particles accumulate near the equator or near the poles, but nothing has been said about the geometric properties of their trajectories. {The strategy we follow to know whether or not the trajectories become planar consists in tracking at each position ${\bf x}$ of the trajectory, the torsion

\begin{equation}
    \tau = \frac{(\dot{\bf x}\times\ddot{\bf x})\cdot\dddot{\bf x}}{|\dot{\bf x}\times\ddot{\bf x}|^2}.
\end{equation}

\noindent We then record the values of $\tau$ in a histogram 
at initial time and after $20\tau_s$ that we show in Fig. \ref{fig:torsion_core_domin} for the monopole and dipole dominating configurations.}

The results indicate that the trajectories of test particles, due to the randomness of the initial conditions, start with a rather wide distribution of values of $\tau$. However during the evolution, torsion tends to small values, which means that the gravitational potential due to the multi-state configuration ``flattens" the trajectories with a sharp peak near zero.

\section{DIFFERENT INITIAL CONDITIONS}
\label{app:otherAss}

\begin{figure*}
\centering
\includegraphics[width=0.24\textwidth]{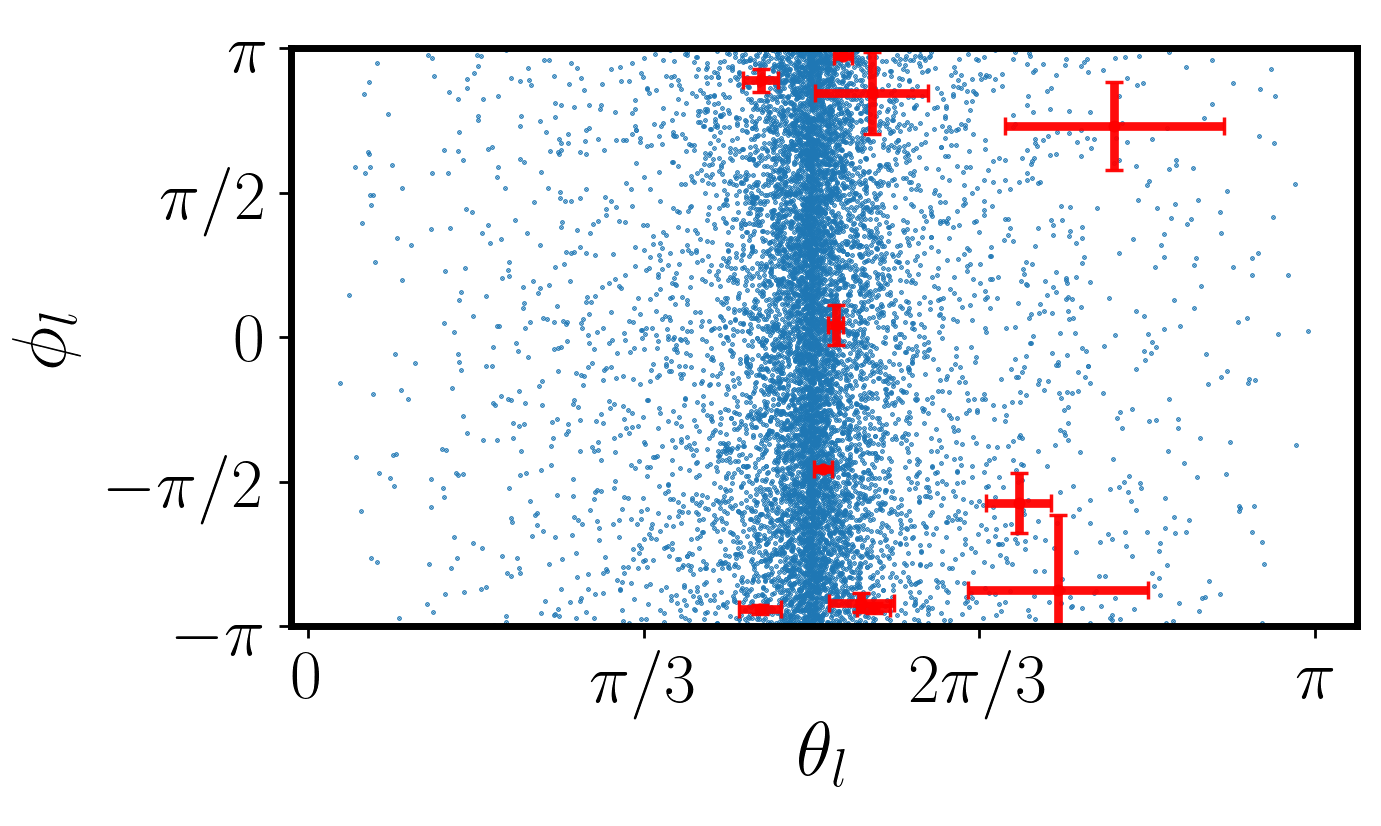}
\includegraphics[width=0.24\textwidth]{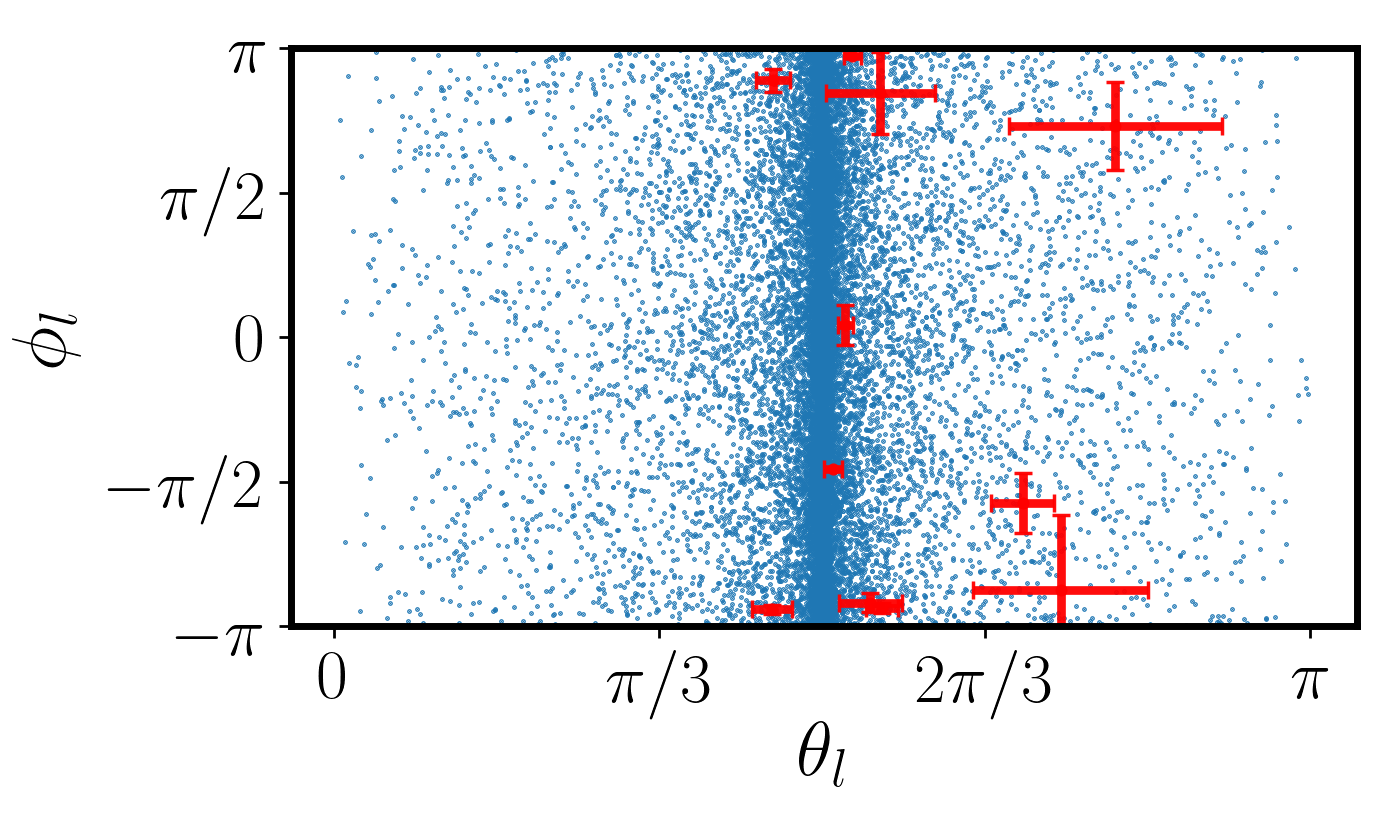}
\includegraphics[width=0.24\textwidth]{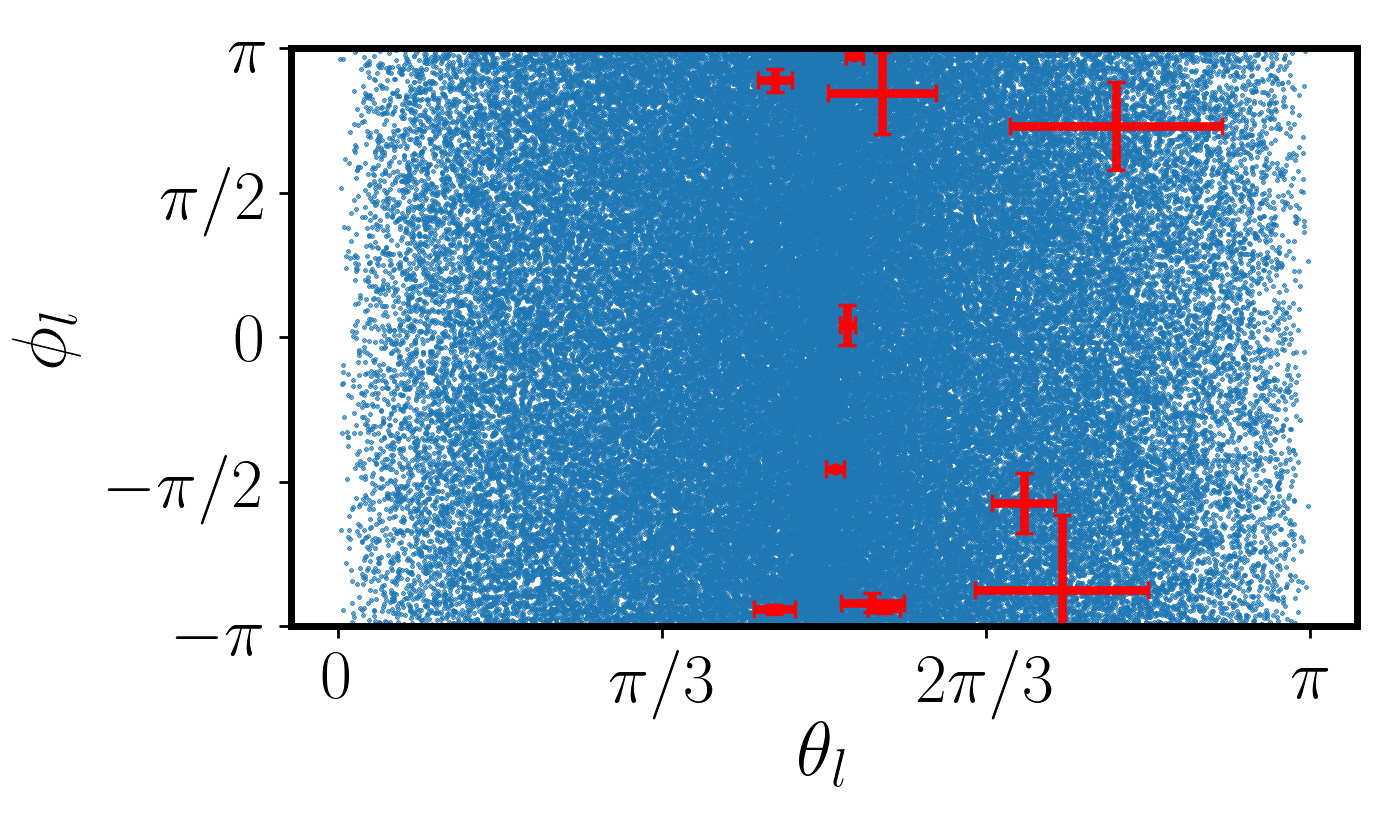}
\includegraphics[width=0.24\textwidth]{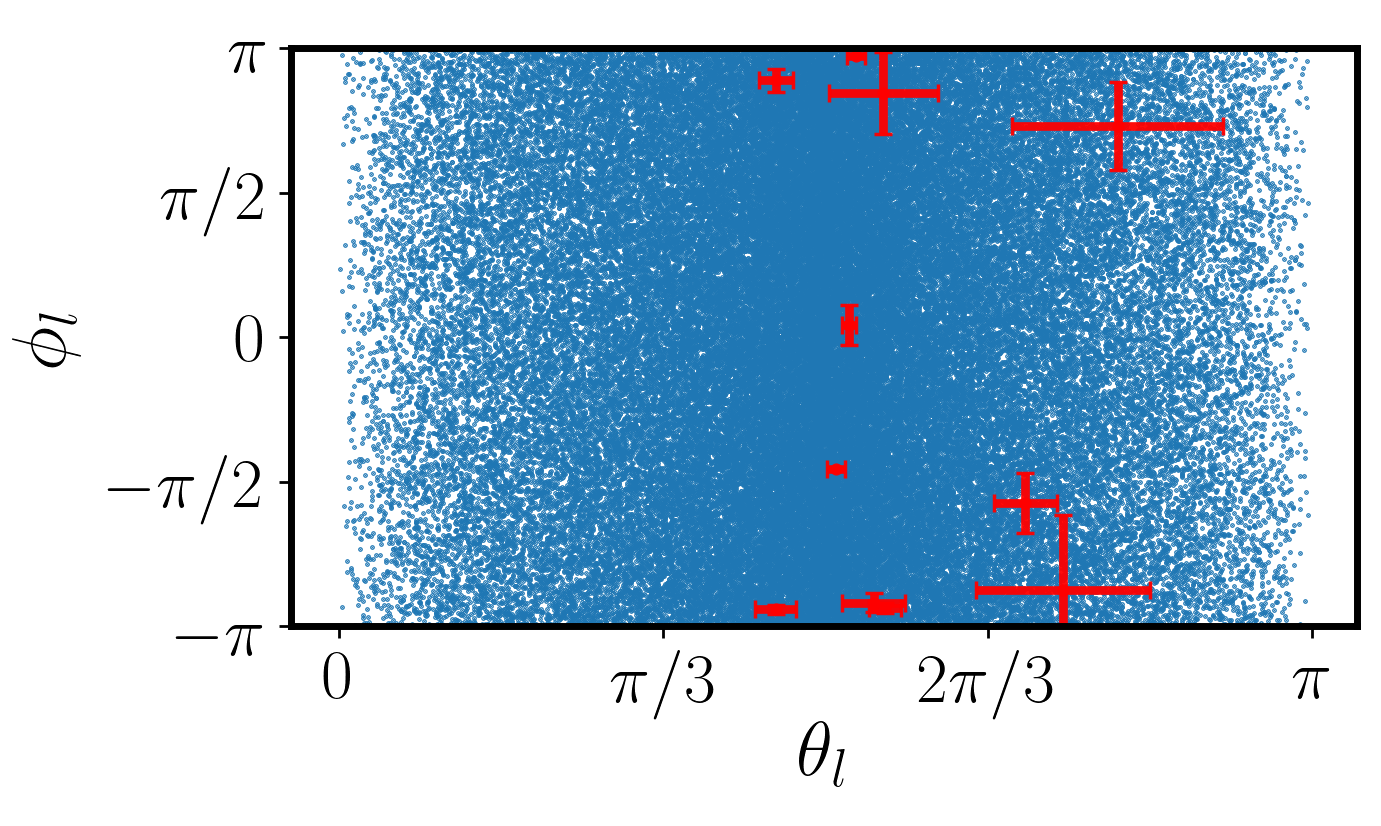}
\includegraphics[width=0.24\textwidth]{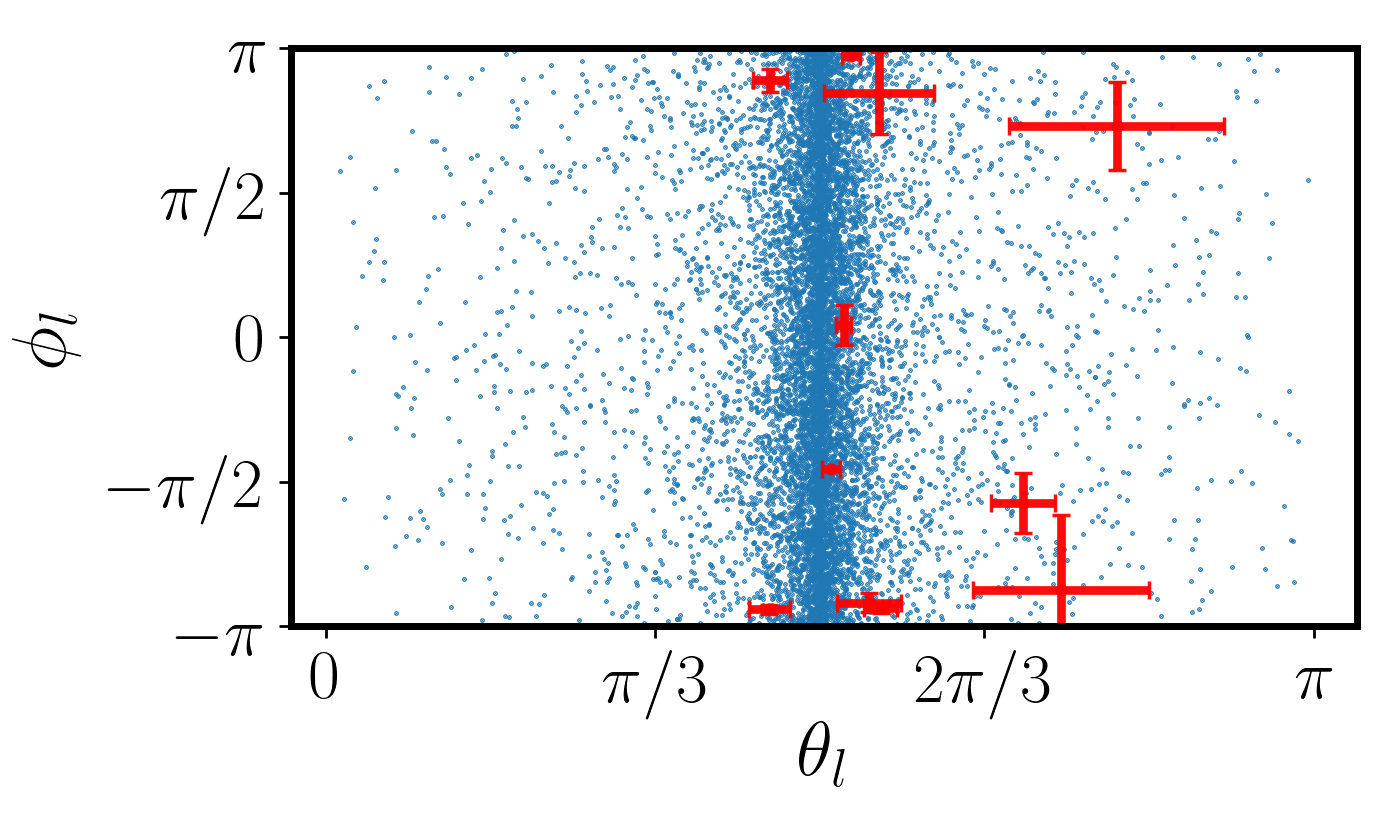}
\includegraphics[width=0.24\textwidth]{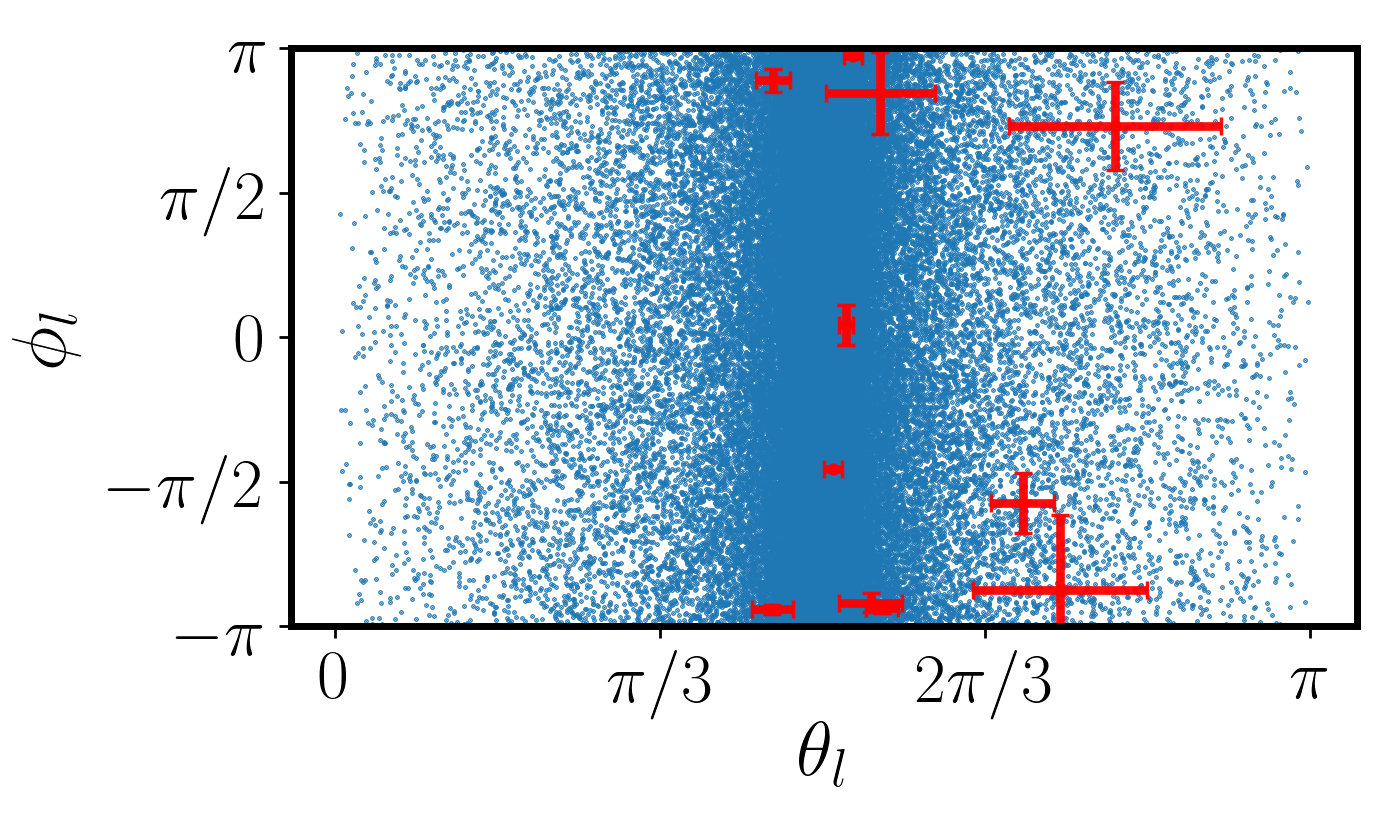}
\includegraphics[width=0.24\textwidth]{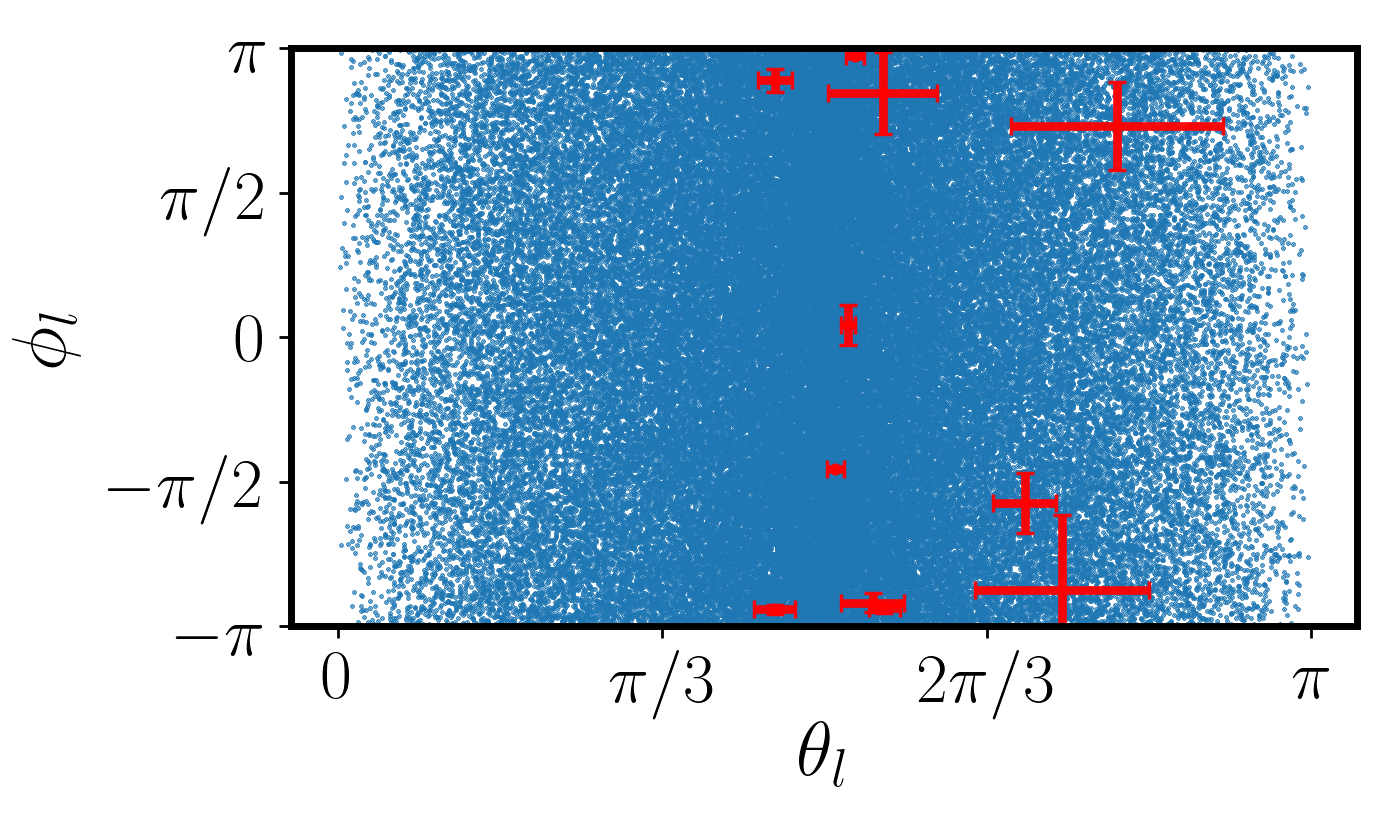}
\includegraphics[width=0.24\textwidth]{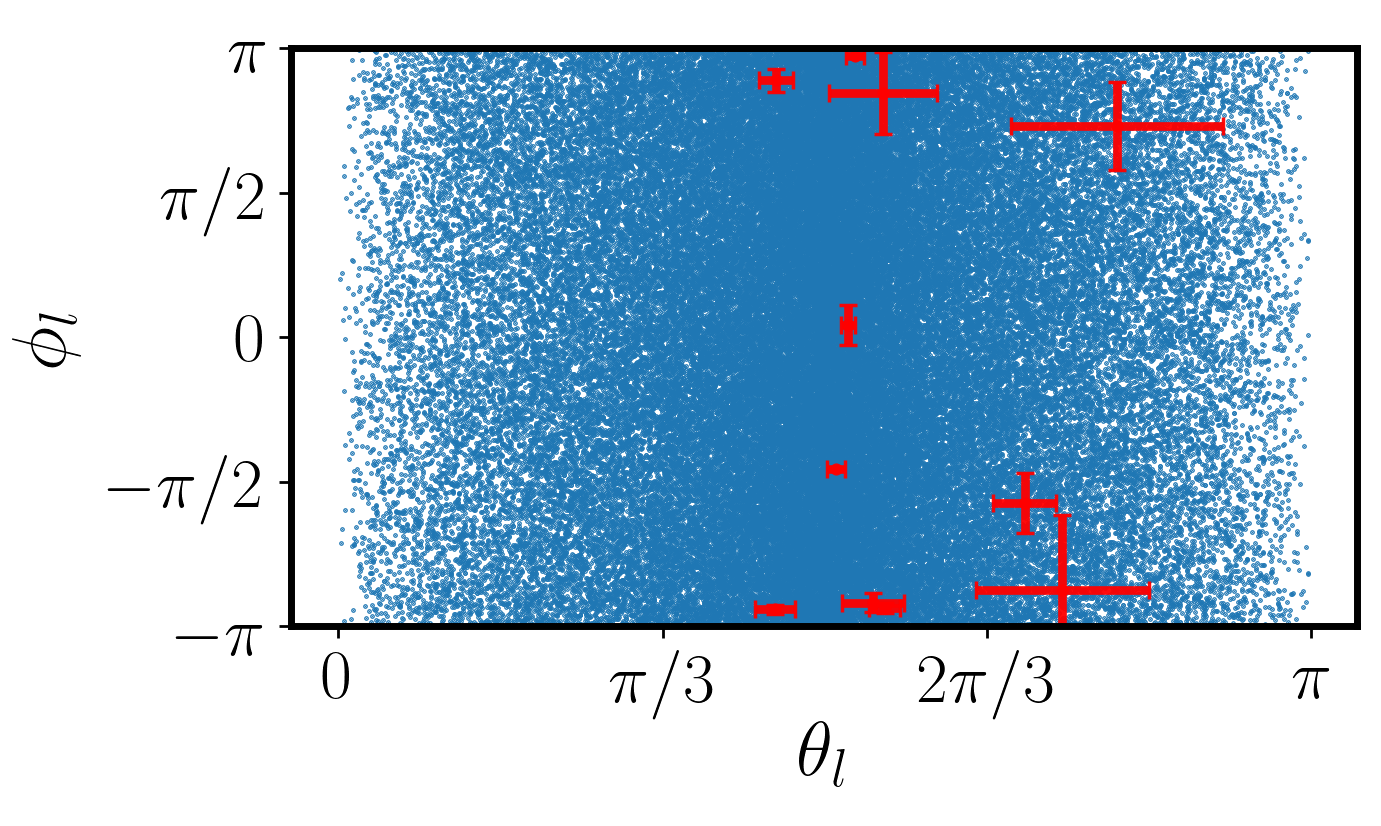}
\caption{{Polar  and azimutal angle of the orbital poles  of  test particles  after $t = 20\tau_s$. From left to right the values of $a$ are 1/4, 1/2, 3/4, 1. In the upper panel the results correspond to the monopole-dominated configuration whereas in the bottom those of the dipole-dominated configuration.}}
\label{fig:polaranglesICs}
\end{figure*}

{At the beginning of Sec. \ref{sec:particles} we described the initial conditions of test particles, specifically that the direction is random whereas the magnitude is bounded by $v_{max}=av_{esc}$. As mentioned, the results presented in the body of the paper correspond to the case $a=1/2$. In this appendix we show the implications of using different values of $a$, specifically 1/4,3/4 and 1, in addition to the case analyzed in depth $a=1/2$.

The results for the orbital poles in each case are shown in Fig. \ref{fig:polaranglesICs} for the two multi-state configurations considered in the paper. Notice that for particles with velocities bounded to be small ($a=1/4,1/2$), the orbital poles show a clear accumulation around $\theta \sim \pi/2$, whereas for the case of fast particles ($a=3/4,1$), the orbital poles distribute nearly isotropically. The reason is that particles are allowed to travel very far away from the influence of the dipolar contribution, from where the configuration looks spherical. 

These results indicate the dependency of the anisotropy in the orbital poles, on the distribution of velocities of test particles. This adds an extra parameter to the analysis of specific galaxy observations.
}

\section{NFW POTENTIAL}
\label{appendixNFW}

{In order to compare the SFDM halo with a CDM one, here we study the effects of a nonspherically symmetric NFW halo whose mass density could resemble that of a multistate SFDM halo, on the same test particles of our analysis. For this we consider a triaxial NFW halo with density profile \citep{triaxialNFW}}
\begin{equation}
    \rho(\varrho) = \frac{\delta_c \rho_{c}}{(\varrho/r_s) (1 + \varrho/r_s)^2}
\end{equation}
{where $r_s$ is the scale length, $\delta_c$ the density contrast, $\rho_c$ the critical density of the universe and}
\begin{equation}
    \varrho^2 = \alpha^2\left( \frac{x^2}{\alpha^2} + \frac{y^2}{\beta^2} + \frac{z^2}{\delta^2} \right). 
\end{equation}
{We distorted along only two directions, so that the triaxiality shows an important dipolar contribution $\alpha = \beta = 0.5$ and $\delta = 1$.} 

\begin{figure}
\centering
\includegraphics[width=8cm]{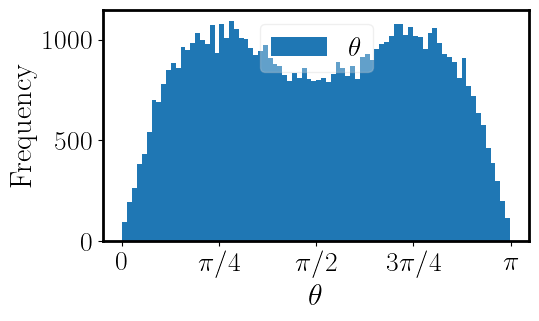}
\caption{{Histogram of test particles after $20\tau_s$ within 30 and 300 kpc for the triaxial NFW configuration.}}
\label{fig:histogramNFW}
\end{figure}

\begin{figure}
\centering
\includegraphics[width=8cm]{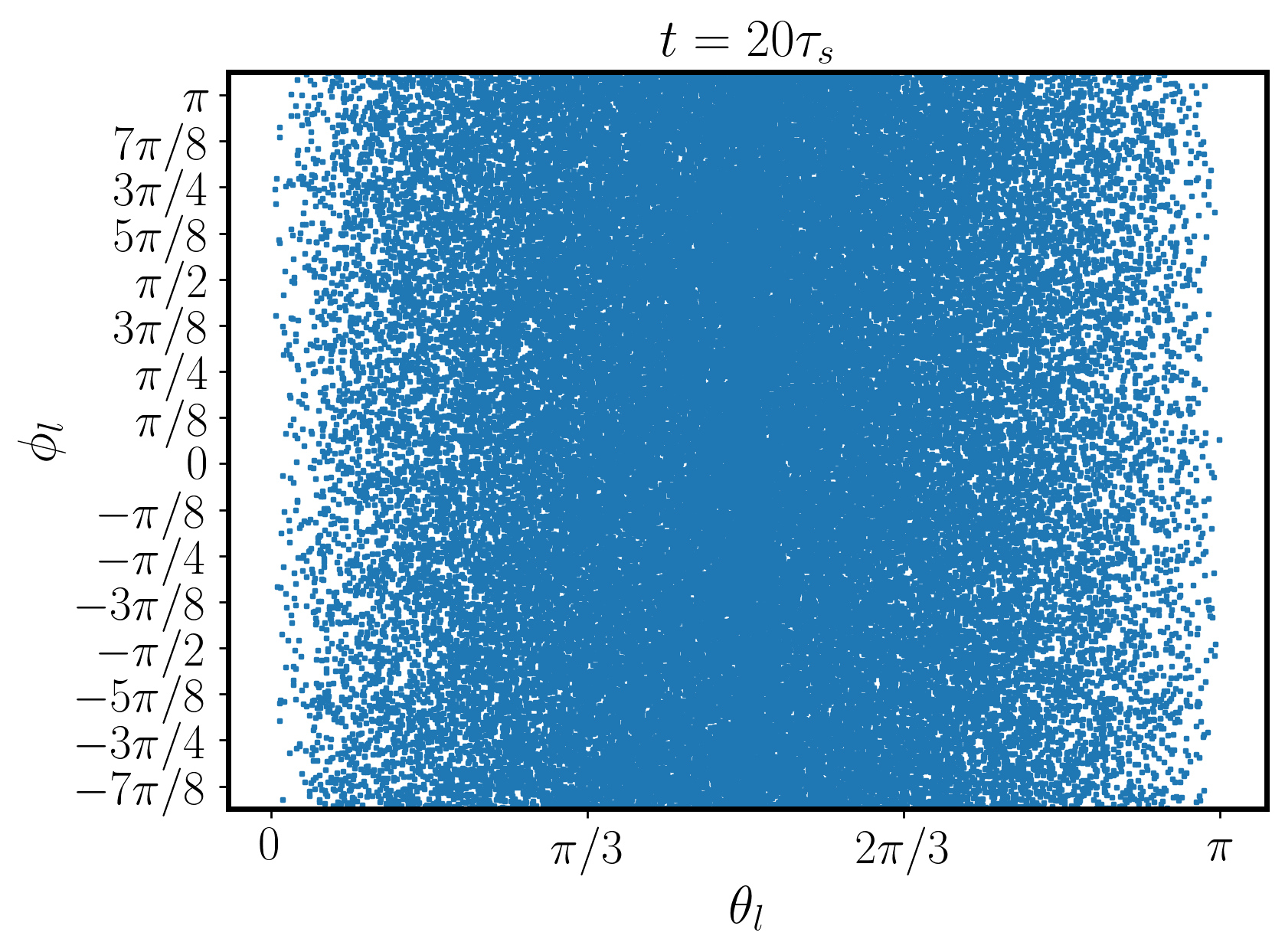}
\caption{{Orbital poles of $10^5$ test particles for the triaxial NFW configuration after $20\tau_s$, unlike the multistate SFDM case, the angular poles remain uniformly distributed.}}
\label{fig:orbital_polesNFW}
\end{figure}

{As in the multistate SFDM configurations, we ran a simulation with $10^5$ test particles with random initial positions and random initial velocities with $a=1/2$,  and found that these accumulate on trajectories near the poles as shown in the histogram of Fig. \ref{fig:histogramNFW}. Unlike the multistate SFDM case, the orbital poles do not concentrate near $\pi/2$, instead they appear isotropically distributed. The distribution of orbital poles for this NFW halo is shown in Fig. \ref{fig:orbital_polesNFW}, which should be compared with Fig. \ref{fig:orbital_poles}. The result is  generic for distorted NFW profiles, since they do  not include the peanut-shape contribution of a (2,1,0) mode as the multistate SFDM halos.}

\bibliography{GAtoms.bib}

\end{document}